\documentclass[10pt, conference, compsocconf]{IEEEtran}
\ifCLASSINFOpdf
\else
\fi

\usepackage{helvet}
\usepackage{courier}

\usepackage{comment}
\usepackage{color}
\usepackage{amsmath}
\usepackage{multirow}
\usepackage{bm}
\usepackage{balance}

\usepackage{times}
\usepackage{graphicx} 
\usepackage{subfigure}

\usepackage{textcomp}
\usepackage{amssymb}
\usepackage{epsfig}
\usepackage{graphicx,subfigure}
\usepackage{enumerate}
\usepackage{amsmath}
\usepackage{amscd}
\usepackage{rotating}
\usepackage{algorithm,algorithmic}
\usepackage{multirow}
\usepackage{url}

\begin{document}
%
\title{Expenditure Aware Rating Prediction for Recommendation}




%
\author{\IEEEauthorblockN{Chuan Shi\IEEEauthorrefmark{1},
Bowei He\IEEEauthorrefmark{1},
Menghao Zhang\IEEEauthorrefmark{1},
Fuzheng Zhuang\IEEEauthorrefmark{2} and
Philip S.Yu\IEEEauthorrefmark{3}}
\IEEEauthorblockA{\IEEEauthorrefmark{1}Beijing University of Posts and Telecommunications, Beijing, China\\Email: shichuan@bupt.edu.cn, hebowei1234@gmail.com, Jack@bupt.edu.cn}
\IEEEauthorblockA{\IEEEauthorrefmark{2}Institute of Computing Technology, CAS, Beijing, China\\Email: zhuangfz@ics.ict.ac.cn}
\IEEEauthorblockA{\IEEEauthorrefmark{3}University of Illinois at Chicago, IL, USA\\Email: psyu@cs.uic.edu}}


\maketitle

\begin{abstract}
The rating score prediction is widely studied in recommender system, which predicts the rating scores of users on items through making use of the user-item interaction information. Besides the rating information between users and items, lots of additional information have been employed to promote recommendations, such as social relation and geographic location. Expenditure information on each transaction between users and items is widely available on e-commerce websites, often appearing next to the rating information, while there is seldom study on the correlation between expenditures and rating scores. In this paper, we first study their correlations in real data sets and propose the expenditure aware rating prediction problem. From the data sets crawled from a well-known social media platform Dianping in China, we find some insightful correlations between expenditures and rating scores: 1) transactions or experiences with higher expenditures usually lead to higher rating scores; 2) when the real expenditures are higher than users' normal spending behavior, the users usually give higher scores; and 3) there are multiple grades of expenditure behaviors. Based on these three observations, we propose an \textbf{E}xpenditure \textbf{A}ware \textbf{R}ating \textbf{P}rediction method (\textbf{EARP}), based on low-rank matrix factorization, to effectively incorporate the expenditure information. Extensive experiments on five real data sets show that EARP not only always outperforms other state-of-the-art baselines but also discovers the latent characteristics of users and businesses.
\end{abstract}

\begin{IEEEkeywords}
Recommender system; expenditure information; matrix factorization;

\end{IEEEkeywords}

%
\IEEEpeerreviewmaketitle

\section{Introduction}
Recommendation has been widely used in e-commerce and becomes a basic tool in WWW. A basic problem in a  recommender system is to predict the rating scores of users on items. Many recommendation methods~\cite{MNIH2007,LIAN2014,FU2014} have been proposed. In order to make recommendations in different applications, many methods use different kinds of information to improve recommendations, besides the rating score information between users and items. With the prevalence of social media, more and more studies devote to social recommender systems, which utilize social relations among users~\cite{MA2008,MA2009}. Location based recommendations are also popular in recent years~\cite{LIAN2014,LIU2013}, which utilize the geographical information to improve the recommendation performances. Other information has also been considered to improve recommendations, such as sentiments~\cite{DIAO2014} and cost~\cite{GE2011}.

\begin{figure}
\centering
  \includegraphics[width=7.5cm]{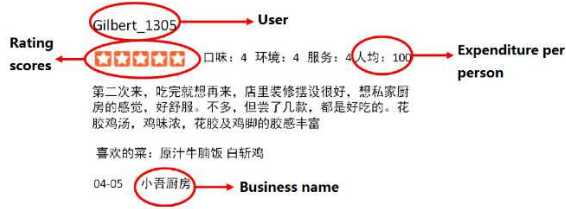}
\caption{An example of expenditure information on Dianping website.}\label{fig:fft:DP}
\end{figure}

In many real world applications, especially e-commerce websites, the interactions between users and items include rating scores, as well as expenditure information on the transaction. For example, on Dianping website\footnote{http://www.dianping.com/} , a well-known social media platform providing reviews of users on businesses in China, users not only rate a business but also provide the expenditure information for the meal at that business. Fig.~\ref{fig:fft:DP} shows an example from Dianping website. When a user receives a service from a business, he will provide the rating score (represented by stars in Fig.~\ref{fig:fft:DP}) as well as the expenditure on the business. Note that the transaction expenditure information is a type of interaction information between users and businesses, not the feature information of either users or businesses. This kind of transaction expenditure information has widely existed, often appearing next to the rating scores, in many real applications. Moreover, there may be correlations between expenditure information and rating scores, and these correlations may help to improve recommendations. However, the effect of expenditure on rating score has not been explored until now.

In this paper, we have crawled review data from Dianping website, where the review information includes a rating score (ranging from 1 to 5) and an average expenditure per person in the transaction\footnote{For example, if 5 persons spend 250  RMB in a restaurant, the average expenditure per person in this transaction is 50 RMB.}. For simplicity, we abbreviate the average expenditure per person in a transaction as expenditure in the following sections. Through analyzing the characteristics of Dianping data sets, we find some insightful correlations between rating scores and expenditures. That is, higher expenditures on businesses usually lead to higher rating scores, and a user tends to give higher scores when his actual expenditure is larger than his normal spending behavior. Moreover, the distribution of expenditures does not follow uniform-distribution, which has multiple expenditure grades satisfying mixture Gaussian distributions. These correlations infer that the expenditure information has the potential to improve the rating score prediction.

However, to integrate the expenditure information into the rating prediction, we face several challenges.
\begin {itemize}
\item Scale mismatch between rating and expenditure. The rating scores are a couple of integer values, while the expenditures are continuous values with a wide range. Moreover, different from rating scores, the same expenditure difference has not a uniform impact at different expenditure grades. For example, RMB 50 is a significant difference for the low-grade expenditure, while it is a trivial difference for the high-grade expenditure.
\item Sparsity with noise on expenditure information. Another challenge lies in that the expenditure information is very sparse, and full of noise. Some users do not mark expenditures or randomly mark expenditures in real applications.
\item More importantly, different individual spending behaviors and distinct business pricing models. Different people have different average expenditure levels. As we shall see, it means a given price range has distinct meaning to different individuals resulting in different expectation or satisfaction levels. For example, an individual with an average expenditure level of RMB 20 will view RMB 50 as  a luxury expenditure, while another individual with an average expenditure level of RMB 100 will view it otherwise. In addition, different businesses have different business models and operate on different price grades. In fact, each business can operate on multiple price grades as we shall see later from the data.
\end {itemize}

In order to effectively utilize the expenditure information and solve the above challenges, we propose an \textbf{E}xpenditure \textbf{A}ware \textbf{R}ating \textbf{P}rediction method (\textbf{EARP}) with a novel optimization objective in which an expenditure term is inferred from expenditure information. According to the observations in real data, we assume that rating scores have a linear correlation with expenditures, and the expenditures can be divided into multiple grades satisfying mixture Gaussian distributions. And thus different users have different sentiments to expenditure grades, and different businesses have different positionings to these expenditure grades. Along this line, we design the expenditure term, which includes the sentiment matrix $W$ and positioning matrix $D$ to represent the sentiment of users and the positioning of businesses on different expenditure grades. Moreover, the expenditure term is used as a correction to the  factorization of rating matrix, which constitutes a new optimization objective under the low-rank matrix factorization. Experiments on five real data sets validate its effectiveness.  This work has the following three contributions.
\begin {itemize}
\item In this work, we systemically study the expenditure aware rating prediction problem, which is to predict the rating score of a user on a business based on historical rating and expenditure information. Through analyzing real data sets with expenditures, we find some insightful correlations between scores and expenditures. These correlations have seldom been studied before, while they are really helpful to improve recommendations.
\item  We propose an EARP method to solve the expenditure aware rating prediction problem. In order to effectively integrate the correlations of rating scores and expenditures, we devise an expenditure term as a correction to the factorization of rating matrix under low-rank matrix factorization, instead of directly factorizing expenditure matrix or simply treating expenditure as features. The expenditure term not only subtly reveals different individual spending behaviors and distinct business pricing models but also effectively alleviates the effect of missing and noise expenditure information. In addition, the one-dimensional clustering on expenditure information with mixture Gaussian model addresses the scale mismatch between rating and expenditure.
\item Extensive experiments validate the effectiveness of EARP. The comparison experiments on five real data sets show that EARP significantly outperforms other state-of-the-art methods and subtly reveals characteristics of users and businesses.
\end {itemize}

\begin {table}[t]
\centering  \small
\caption { Statistics of five data sets}\label{tab:aTable}
\resizebox{0.48\textwidth}{!}
{
\begin{tabular}{|c|c|c|c|c|c|}
\hline
    &BJ&SH&GZ&HZ&CD \\ \hline
\# Users&176k&314k&73k&103k&53k \\ \hline
\# Businesses&194k&205k&167k&151k&148k \\ \hline
\# Reviews&2335k&3956k&1114k&947k&696k \\ \hline
Sparsity ratio &0.07\textperthousand&0.06\textperthousand&0.09\textperthousand&0.05\textperthousand&0.09\textperthousand\\ \hline
Ave. Rating&3.78&3.77&3.72&3.64&3.68 \\ \hline
Ave. Expenditure&70.00&75.96&68.50&61.63&63.71 \\ \hline
Ave. \# Rev. of User&13.20&12.59&15.07&9.16&13.09 \\ \hline
Ave. \# Rev. of Bus.&11.98&19.27&6.64&6.26&4.68\\ \hline

\end{tabular}
}
\end{table}

\begin{figure*}
\centering
\subfigure[Score dist. on BJ]{\label{fig:fft:sdd1}
  \includegraphics[width=4.0cm]{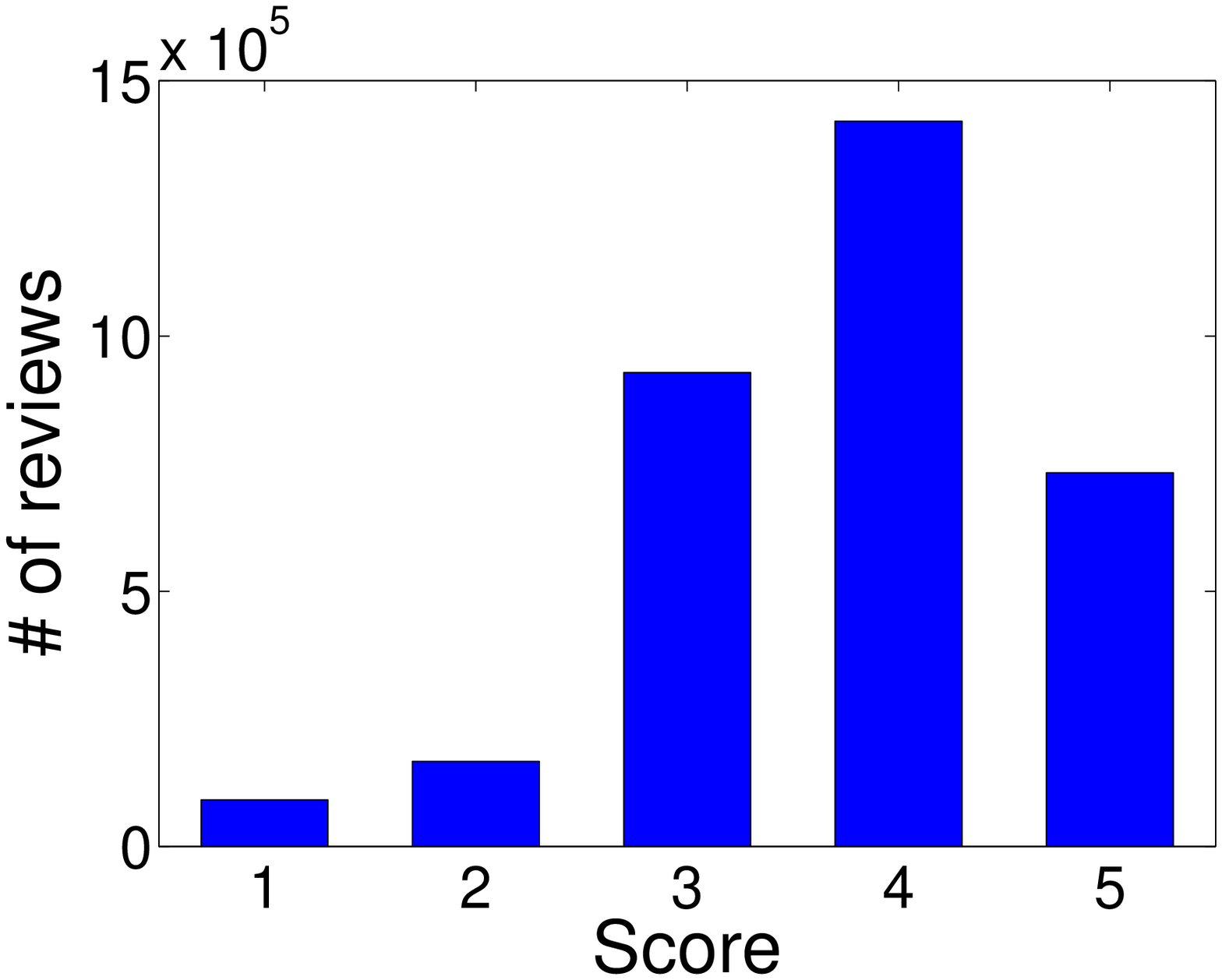}
}
\subfigure[Score dist. on SH]{\label{fig:fft:sdd2}
  \includegraphics[width=4.0cm]{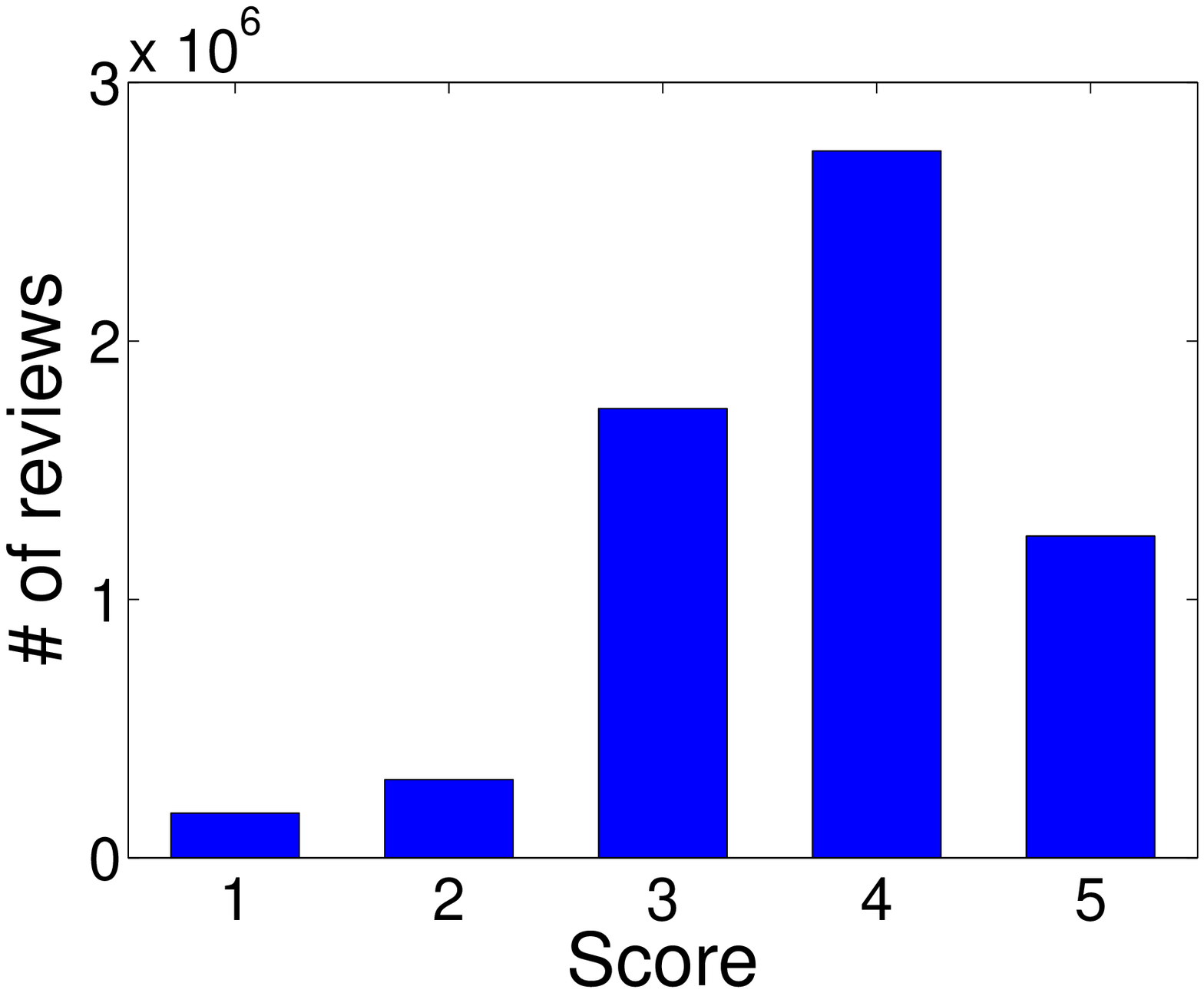}
}
\subfigure[Expenditure dist. on BJ]{\label{fig:fft:pdd1}
  \includegraphics[width=4.0cm]{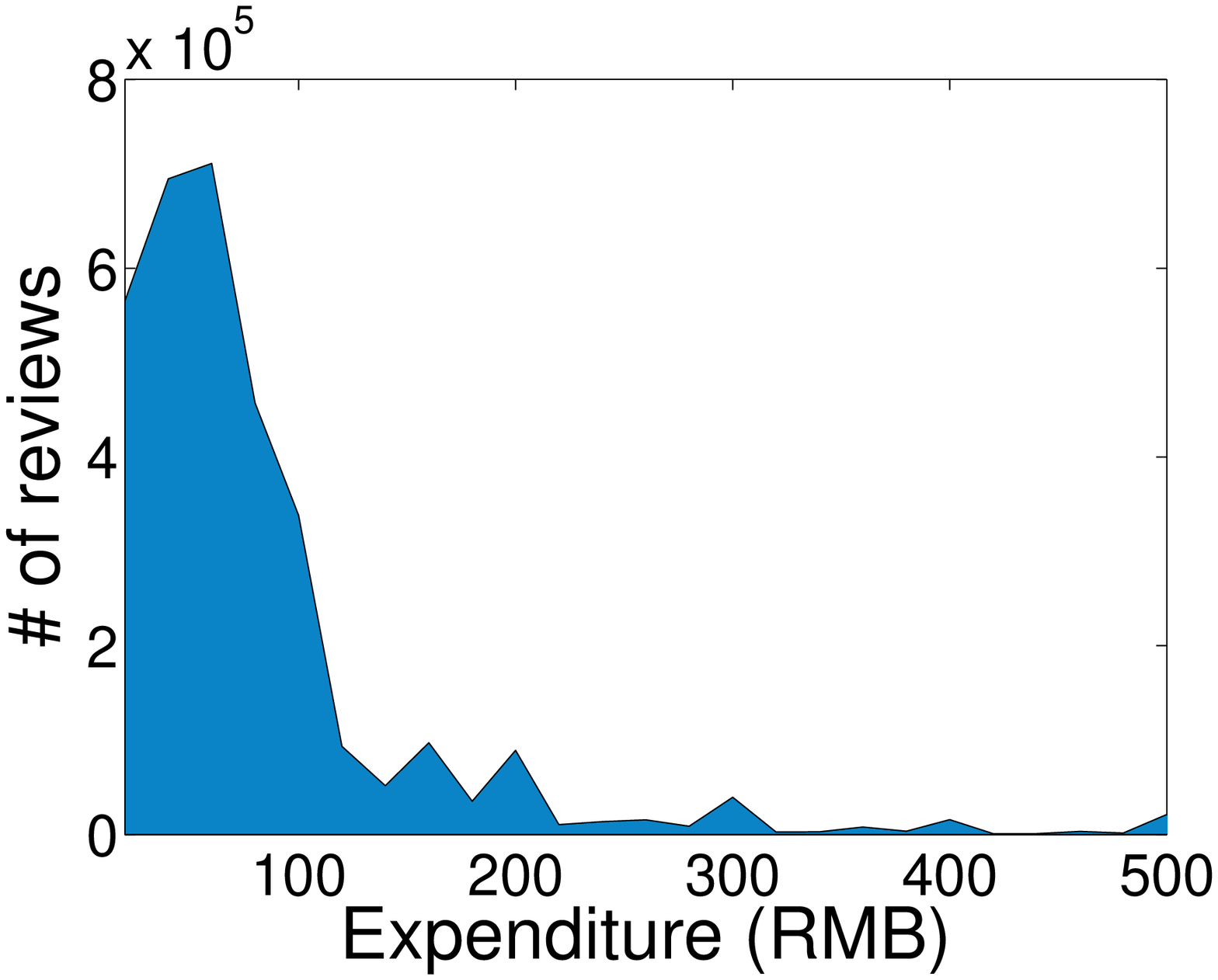}
}
\subfigure[Expenditure dist. on SH]{\label{fig:fft:pdd2}
  \includegraphics[width=4.0cm]{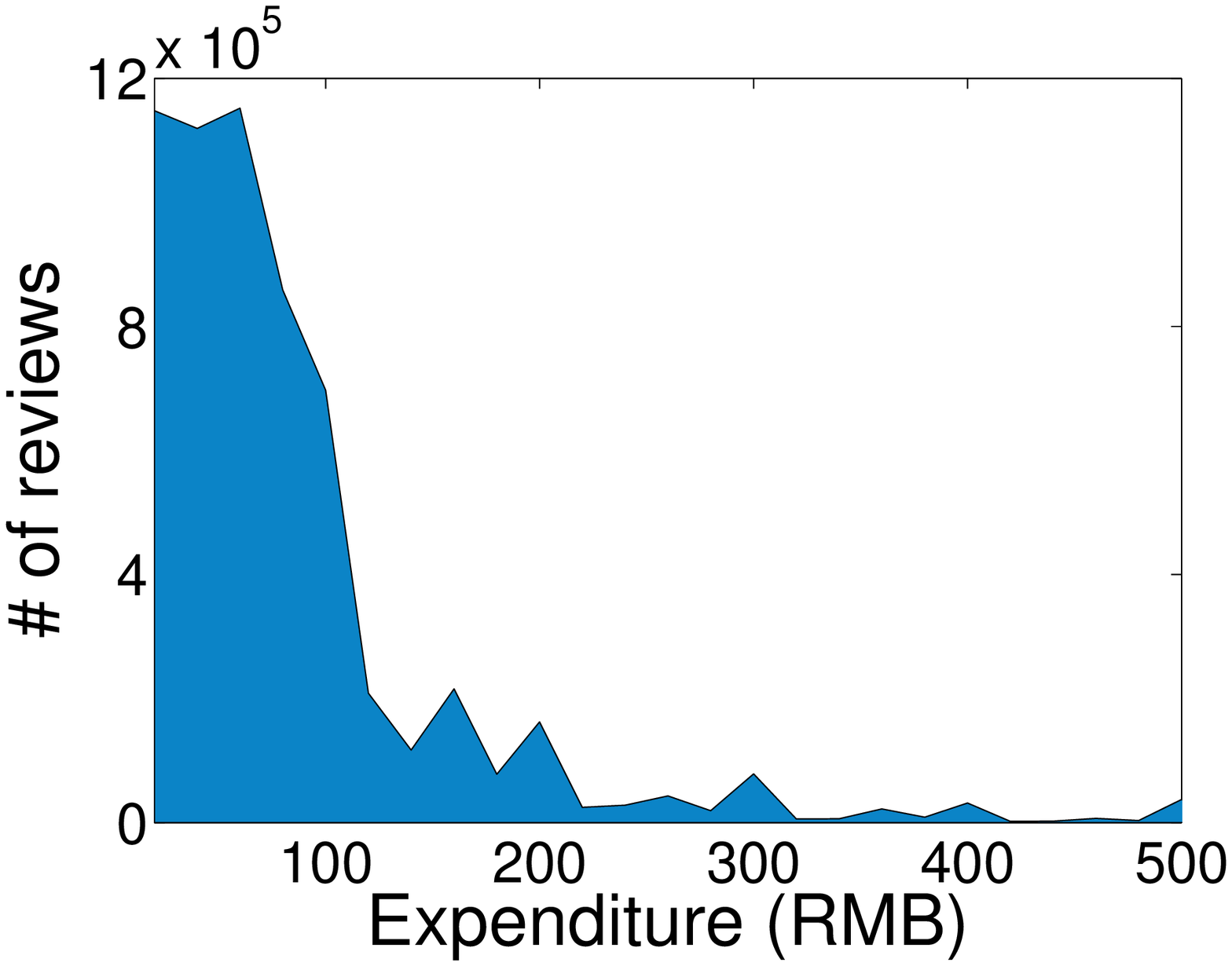}
}
\caption{The distributions of rating scores and expenditures on reviews. }\label{fig:fft:dis}
\end{figure*}

\section{Data Preparation and Observations}
\subsection {Data Preparation}
We crawled data from a well-known social media website Dianping, which provides a review platform for businesses and entertainments. On Dianping website, a user can give a review to a business when he has received a service in this business. The review information includes a rating score (ranging from 1 to 5) and an average expenditure for each person in this transaction. We independently sample a set of businesses and users in five representative big cities, Beijing, Shanghai, Guangzhou, Hangzhou, and Chengdu (Shortly referred as BJ, SH, GZ, HZ, CD)~\footnote{We are working on getting approval from Dianping to make an anonymized version of the dataset available.}. And we get the review information, including rating score ${R}\in\mathbb{R}{^{m\times n}}$ and expenditure \(C\in\mathbb{R}{^{m\times n}}\), between users and businesses, where $m$ is the number of users and $n$ is the number of businesses. Moreover, the $R_{i,j}$ and $C_{i,j}$ represent the rating score and average expenditure of user $i$ on business $j$, respectively.

Since the transactions with very high expenditures (e.g., larger than RMB 1000) are very rare, we simply delete them. The statistics information of these data sets are summarized in Table~\ref{tab:aTable}. We can find that, although these data sets have different sizes, they have similar average rating scores and average expenditures. Moreover, the review information in these data sets all are very sparse. Due to the similar distributions and very similar analysis results on these five data sets, we only show the results on BJ and SH for saving space in this section.

Furthermore, we compute statistics on the distributions of rating scores and expenditures of users. The results are shown in Fig.~\ref{fig:fft:dis}. We can find that 1) most rating scores range from 3 to 5, and the maximal number of users give the score 4 to businesses. 2) The expenditures have obvious long-tailed distributions, that is, most of them are smaller than RMB 100 , and few larger than RMB 300. 3) We can also find that there are several peaks in Fig.~\ref{fig:fft:pdd1} and Fig.~\ref{fig:fft:pdd2}.

\begin{figure*}
\centering
\subfigure[Relation of expenditures and scores on BJ]{\label{fig:fft:rsd1}
  \includegraphics[width=4cm]{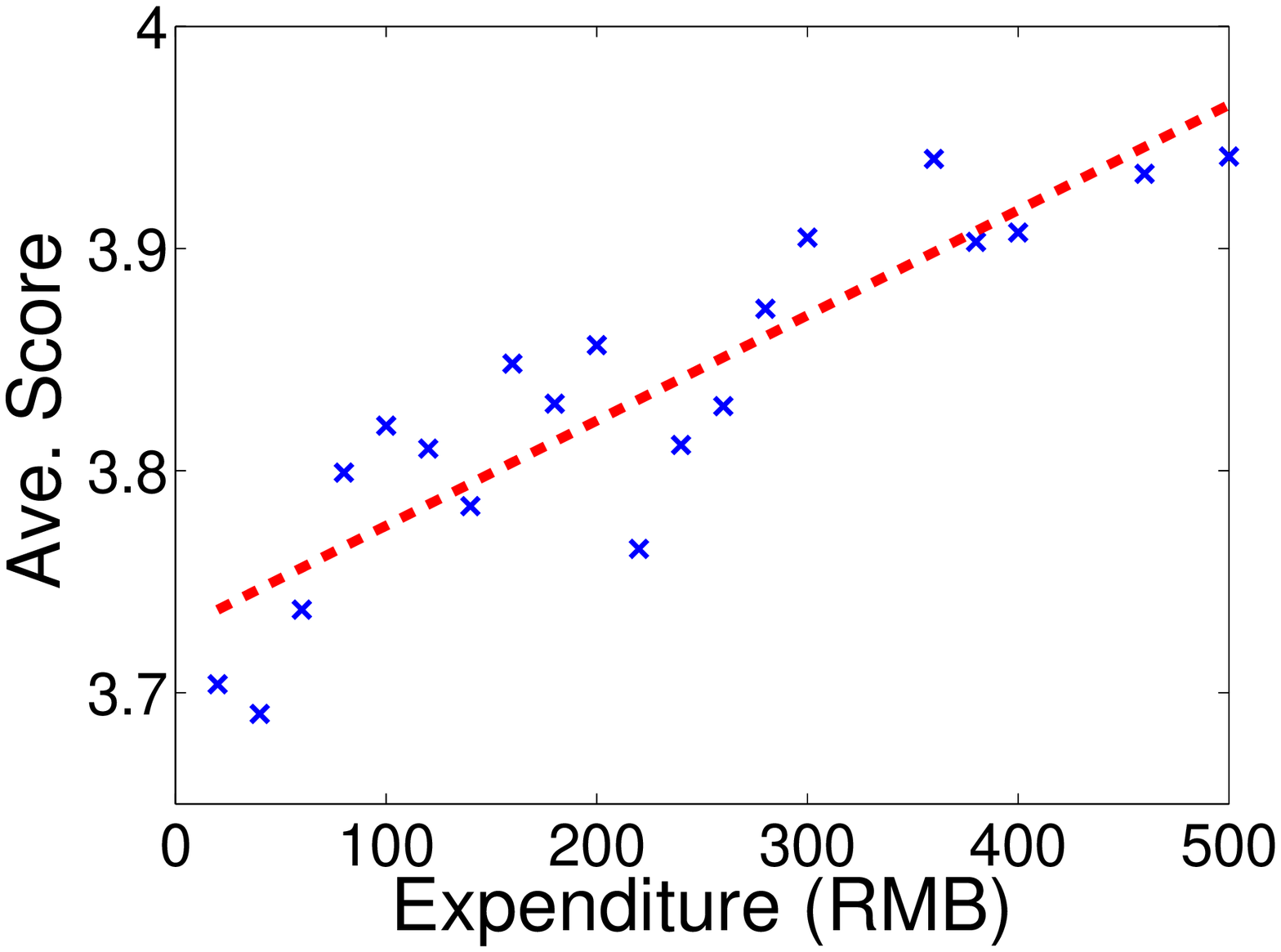}
}
\subfigure[Relation of expenditures and scores on SH]{\label{fig:fft:rsd2}
  \includegraphics[width=4cm]{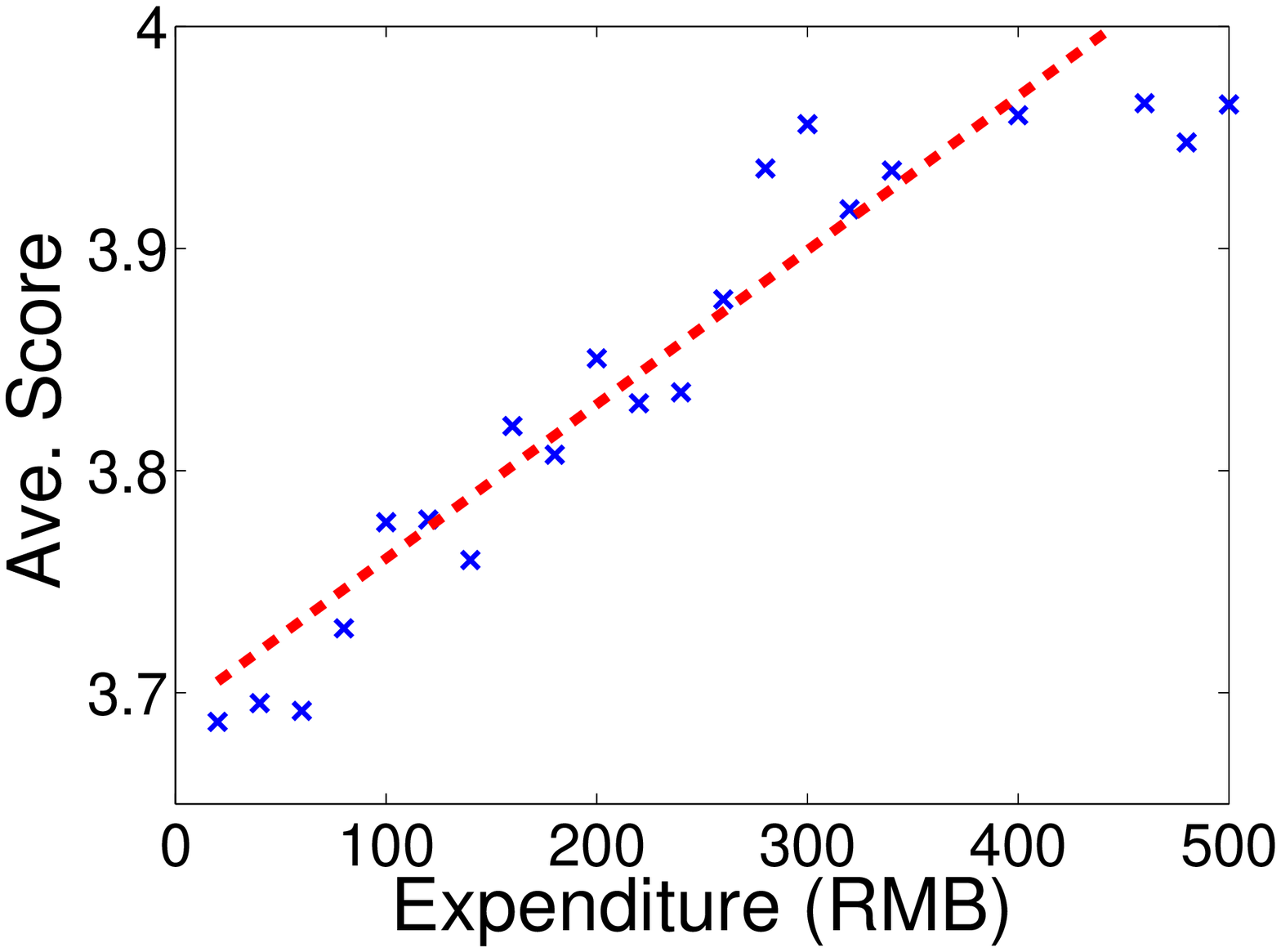}
}
\subfigure[Relation of user spendings and scores on BJ]{\label{fig:fft:rpd1}
  \includegraphics[width=4cm]{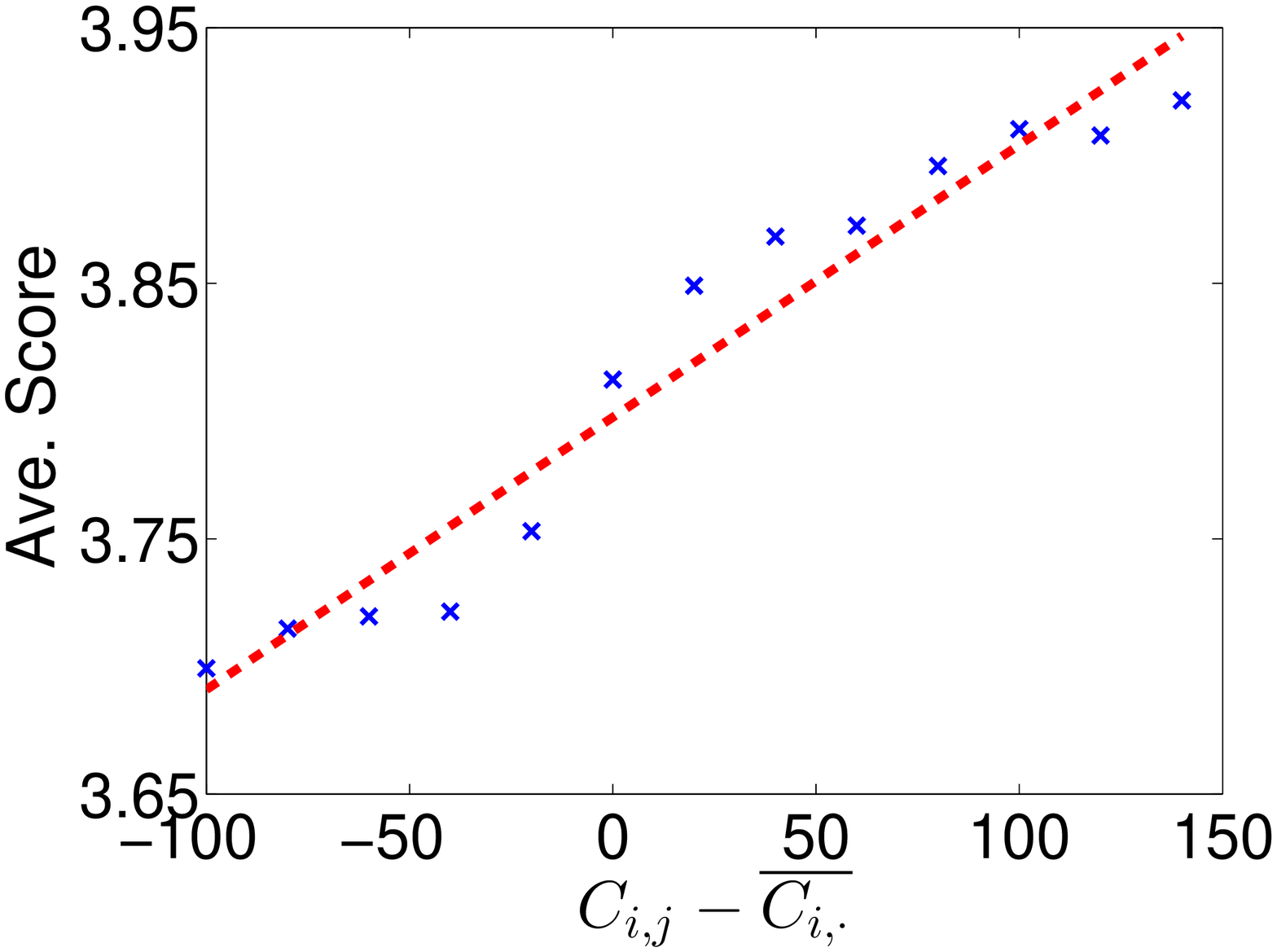}
}
\subfigure[Relation of user spendings and scores on SH]{\label{fig:fft:rpd2}
  \includegraphics[width=4cm]{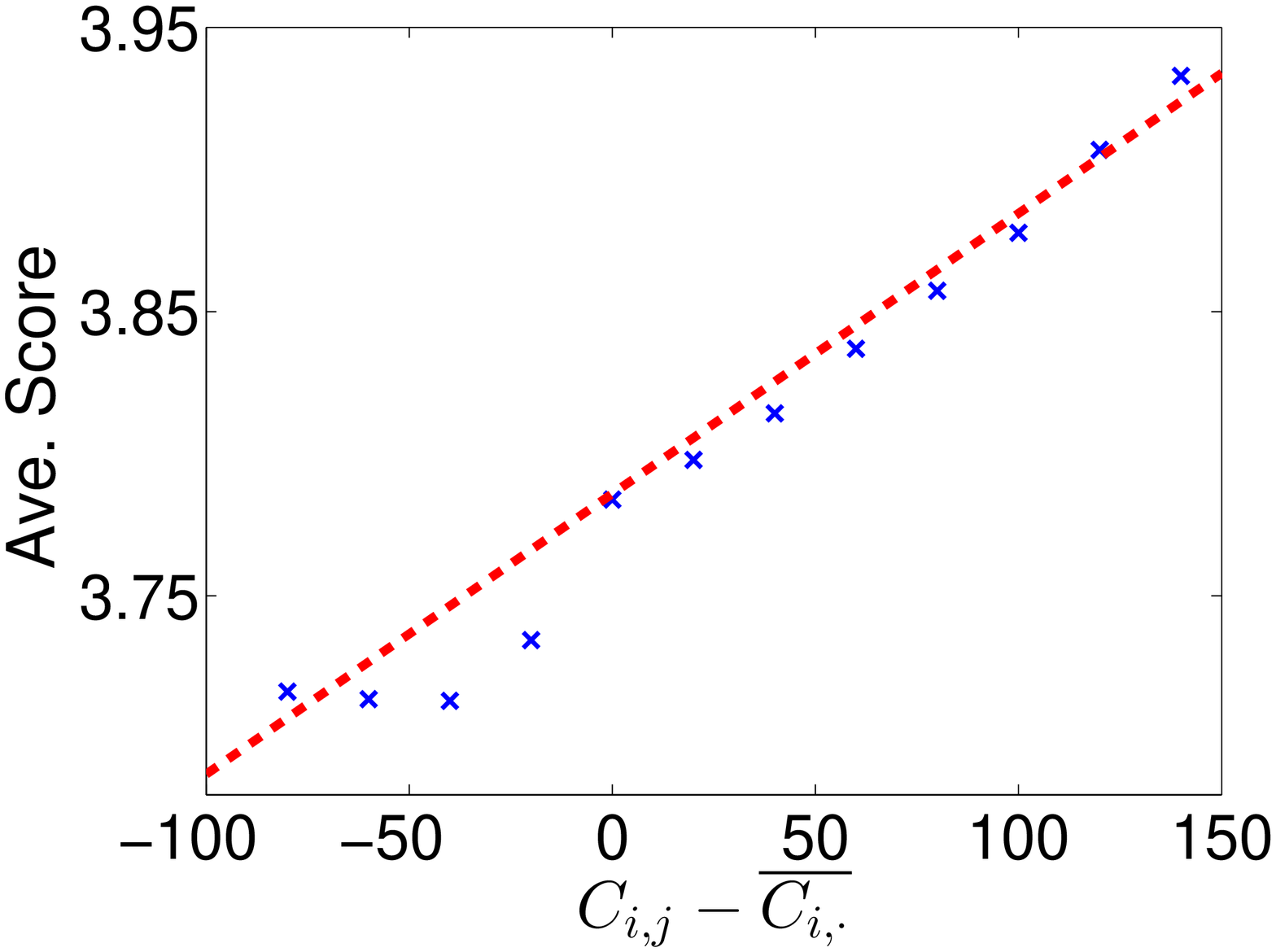}
}
\caption{The effect of expenditures on rating scores.  }\label{fig:fft:relationship}
\end{figure*}

\subsection {Observations} \label{sec:ob}

Since a user may receive services from many businesses and a business also provides services to many users, we first introduce the following two concepts to describe the characteristics of users and businesses.
\begin {itemize}
\item \emph{User spending} is the average expenditure of a user over businesses. We define $\overline{C_{i,\cdot}}$ to represent the user spending of the user $i$, and it can be calculated as follows,
\begin{equation}\label{eq:up}
\overline{C_{i,\cdot}}=\frac{\sum_{j=1}^{n}C_{i,j}}{|\{C_{i,j}|C_{i,j}\neq 0,j=1,\cdots,n\}|}\\.
\end{equation}
\item \emph{Business pricing} is the average expenditure of a business over users. We define $\overline{C_{\cdot,j}}$ to represent the business pricing of the business $j$, which can be calculated as follows,
\begin{equation}\label{eq:up}
\overline{C_{\cdot,j}}=\frac{\sum_{i=1}^{m}C_{i,j}}{|\{C_{i,j}|C_{i,j}\neq 0,i=1,\cdots,m\}|}\\.
\end{equation}
\end{itemize}
We can find that the user spending reflects the consumption habit and ability of a user, and the business pricing embodied the grade and reputation of a business.  Through intensively exploring the characteristics of data, we can observe three interesting phenomena about the correlations of rating scores and expenditures.

\subsubsection{Observation 1: Expenditure correlation rule} \label{sec:ob1}

We discretize the expenditures with an interval of 20, and then calculate the average rating scores in each expenditure range. Each expenditure and its corresponding average rating score on BJ and SH are plotted in Figs.~\ref{fig:fft:rsd1} and ~\ref{fig:fft:rsd2}. Thus, these results can indicate the relation between rating scores and expenditures.

It is obvious that \emph{the rating scores are positively correlated to expenditures}. That is, if a user has a high expenditure on a business, he is likely to give high rating score to the business. This observation is seldom noticed, while it is reasonable. When a user has a high expenditure on a business, he usually gets better services and enjoys the business, so he is more likely to give a higher rating score to the business.

\subsubsection{Observation 2: User spending correlation rule} \label{sec:ob2}
First, we calculate the differences between the actual expenditure and the average user spending for each user $i$, i.e., $C_{i,j} - \overline{C_{i,\cdot}}$, and discretize the expenditure differences with an interval of 20. Then we calculate the average rating scores in each expenditure difference range. Each expenditure difference and its corresponding average rating score on BJ and SH are plotted  in Figs.~\ref{fig:fft:rpd1} and ~\ref{fig:fft:rpd2}.

 In this study, we try to find how the differences between the actual expenditure and average user spending affect the rating scores. It is clear that \emph{rating scores are also positively correlated to the differences between the actual expenditure and user spending}. That is, if a user spends more money than his usual spending (i.e., average expenditure) on a business, he tends to give a relatively higher score. This is reasonable, since he generally gets better services in this situation. It also shows that, if two users have the same expenditure on a business, the user with high average spending is likely to give a lower score, while the user with low average spending is likely to give a higher score. This is also natural. The user with high average spending usually has higher requirement and taste, and thus has more complaint in this situation. However, it is not the case for the user with low average spending. As a consequence,  the rating score is correlated to expenditures, as well as user spendings. In fact, we note that relation between spending and happiness has also been actively studied in psychology \cite{matz2016money}. Our observation also confirm this psychological rule.


\subsubsection{Observation 3: Multiple expenditure grades rule} \label{sec:ob3}
As shown in Figs.~\ref{fig:fft:pdd1} and~\ref{fig:fft:pdd2}, there are several peaks of the expenditures, which is also reasonable. That is to say, \emph{the expenditures of users on business are distributed in multiple expenditure grades}. Specifically, for a user with low spending, his main spending is around RMB 50 , while he also sometime spends at RMB 100 or 200. For a user with high spending, his main spendings are round RMB 150 , while he also spends at RMB 50  occasionally. Similarly, the business pricings of businesses also have multiple expenditure grades. For a high-end business, its main spending is around RMB 300 , while the expenditures of some users may be RMB 150  or 500. For a chophouse, its main spending is around RMB 30, while there are also some expenditures around RMB 50.
\section{The EARP Model}
In this section, we first describe our expenditure aware rating prediction problem. And then we propose three versions of EARP model to integrate three rules observed in Section~\ref{sec:ob}. Correspondingly, the three models are called EARP-E, EARP-U, and EARP-M, respectively. In the design of EARP, we propose delicate strategies to address those challenges mentioned above.

\subsection {Problem Definition}
The problem of \emph{expenditure aware rating prediction} is to predict the rating score of a user on an item based on ratings and expenditures information collected from a web site. Similar with traditional recommendation, we have a rating matrix \(R\in\mathbb{R}{^{m\times n}}\) between users and items. In addition, we also have an expenditure matrix \(C\in\mathbb{R}{^{m\times n}}\) to represent transaction expenditure of users  on items. Please note that the expenditure is a type of interaction between users  on items, not the features of users or items.

A direct solution for this problem is to factorize the rating matrix and expenditure matrix simultaneously and share some common latent factors. However, it is not a good solution, because of the challenges mentioned above. For example, the scale mismatch between rating and expenditure makes the factorized latent factors have different scales. Moreover, the sparsity with noise on expenditure information makes the direct factorization of the expenditure matrix introduce much noise, which inevitably hurts recommendation performances. Another optional solution is to simply treat expenditure information as one type of features, while it fails to utilize the rich correlation of expenditures and rating scores. More importantly, these solutions cannot reveal insightful characteristics of users and businesses. The experiments in Section 4.3 (see the CoMF and FM models respectively) also confirm that these solutions have poor performances.

In this paper, we introduce a novel expenditure term as a correction of the factorization on rating matrix under low-rank matrix factorization framework. The expenditure term makes full use of the correlation of expenditures and rating scores. Moreover, the expenditure term employs the probability distribution to represent sentiment of users and positioning of businesses to expenditure grades, which not only eliminates the effect of expenditure scale and noise but also reveals user spending behaviors and business pricing models.


\subsection {Low-Rank Matrix Factorization}
We firstly introduce the basic model to utilize the rating information. Given a rating matrix, one-widely used approach for item recommendation is low-rank matrix factorization ~\cite{MNIH2007}. This method involves mapping users and businesses into a joint latent space with the number of dimensions \(k\ll \min(m,n)\), such that a user's preference for a business modeled as inner product between them in that latent space. The mapping is achieved by approximating the rating matrix by solving the following optimization problem,
 \begin{equation}
 \underset{P,Q}{\min} \quad \frac{1}{2}\parallel I\odot (R-PQ^{\top})\parallel_{F}^{2}+\gamma (\parallel P\parallel_{F}^{2}+\parallel Q \parallel_{F}^{2} ),
  \end{equation}
where $P\in \mathbb{R}{^{m\times k}}$ and $Q\in \mathbb{R}{^{n\times k}}$ are user latent factor and business latent factor, respectively. \( \parallel \cdot \parallel_{F}\) is the Frobenius norm of matrix, $\top$ is the transpose of a matrix or a vector. $\odot$ is the element-wise product of two matrixes. $I$ is the indicator function, i.e., $I_{i,j} = 1$ if $R_{i,j}$ has a rating score, else $I_{i,j} = 0$. In order to reduce the generalization error, regularization terms are usually added to this objective function to shrink the latent vectors toward zero and $\gamma$ represents the regularization parameter.

 \subsection {EARP-E model integrating expenditure correlation rule}
 According to the Observation 1 shown in Section~\ref{sec:ob1}, the rating scores are positively correlated to the expenditures. That is to say, the higher expenditure users paid for a business, the higher rating score they may give. So users with different expenditures have different rating bias. We can utilize the rating bias, reflected by expenditures, to correct the factorization of rating matrix.

Inspired by the utilization of geographic information in \cite{LIAN2014}, we add an expenditure item, reflecting the impact of expenditure to rating scores, as a correction of the rating bias of users. However, this will face with a great difficulty: scale mismatch of rating score and expenditure. Here we design a delicate strategy to make them have the same scales. Specifically, we use the business pricing $\bm{v}$, which represents the average expenditure of a business,  as a component of the expenditure item. The business pricing $\bm{v}$ can be deduced from the expenditure matrix $C$ as follows,
\begin{equation} \label{eq:vj}
\bm{v}_{j}=\overline{C_{\cdot,j}}=\frac{\sum_{i=1}^{m}C_{i,j}}{|\{C_{i,j}|C_{i,j}\neq 0,i=1,\cdots,m\}|}\\.
\end{equation}
Since $\bm{v}_{j}\in[0, 1000]$, so we normalize it into [0, 1] to make the scale more comparable to the ones of rating scores. Because rating scores are more sensitive to low expenditures, we first use the log function and then employ the Min-Max normalization technique to compress the business pricing $\bm{v}$ into [0, 1]. After that, we extend the row vector $\bm{v}^{\top}$ into a $m\times n$ matrix $V$ through  duplicating the row vector $\bm{v}^{\top}$.
A simple EARP-E model integrating expenditure item is to add the expenditure factor in low-rank matrix factorization as the correction of rating bias of users.
\begin{equation}\label{eq:ob1}
\underset{P,Q,w}{\min}\ \frac{1}{2}\parallel I\odot (R-PQ^{\top}-wV)\parallel_{F}^{2}+\gamma (\parallel P\parallel_{F}^{2}+\parallel Q \parallel_{F}^{2} ),
\end{equation}
where $w$ is a weight parameter representing the sentiment on expenditures for all users. And the adjusted weight parameter $w$ can automatically confine the rating scores and expenditures into the same scale. Compared to the basic low-rank matrix factorization, the added expenditure item in EARP-E compels the latent factors $P$ and $Q$ to adjust according to expenditures.

\subsection {EARP-U model integrating user spending correlation rule}
The above model has a fixed weight parameter for all users. However, the user spending correlation rule in Section~\ref{sec:ob2} shows that users with different spending have different effects on rating scores. So we need to design a personalized weight for each user. Here we define a weight vector $\bm{w}\in \mathbb{R}^{m\times 1}$, which represents the personal sentiment on expenditures for each user. That is to say it can capture individual behavior difference or average expenditure level. Then EARP-U model is formulated as
\begin{equation}\label{eq:ob2}
\begin{aligned}
\underset{P,Q,\bm{w}}{\min}&\ \frac{1}{2}\parallel I\odot (R-PQ^{\top}-\bm{w}\bm{v}^{\top}) \parallel_{F}^{2}\\&+\gamma (\parallel P \parallel_{F}^{2}+\parallel Q \parallel_{F}^{2})+\beta \parallel \bm{w} \parallel_{F}^{2},
\end{aligned}
\end{equation}
where $\gamma$ and $\beta$ are the trade-off parameters.

\subsection {EARP-M model integrating multiple expenditure grades rule}
In above models, there is a implicit assumption that there is only one expenditure grade for users and businesses. However, the multiple expenditure grades rule in Section ~\ref{sec:ob3} shows that a user may make consumption on multiple expenditure grades and the business pricing of a business also may cover a wide expenditure range. Moreover, the same expenditure difference may have totally different impacts at different expenditure grades. For example, there are no significant difference between RMB 300  and 350 , while the difference is significant between RMB 50  and 100. RMB 50 means low consumption, while RMB 100  may mean moderate consumption. So we need to divide the continuous expenditures into different grades.

The next step is how to divide the expenditures into different grades. A naive idea is to divide the expenditures into several grades with a fixed interval. For example, the expenditures are divided into (0,50], (50,100], (100,150], and so on. But this method is not discriminative. For example, the snack shop may cost only RMB 10, while the common restaurant may cost RMB 50. This method may wrongly put them into the same grade.

In fact, we can clearly observe that there are several peaks of the expenditures in Figs. 2(c) and 2(d), so we consider the division of expenditures as one dimensional clustering problem, and assume that these expenditures are generated by mixtures of Gaussian models. Given the expenditure matrix $C$, we can get the means and variances of mixture Gaussian models with the Expectation-Maximization algorithm~\cite{Dempster77maximumlikelihood}. The detailed EM algorithm for our data sets is shown in Algorithm 1. Each Gaussian model can represent an expenditure grade. $T$ is the number of Gaussian models, i.e., the number of expenditure grades.  Then, each business has an positioning to these expenditure grades. Here we define $D\in \mathbb{R}{^{n\times T}}$ to represent the positioning of business to expenditure grades, which reflects  business pricing models. From the EM algorithm, we can get $T$ Gaussian models  $G=(N(\bm{\mu},\Sigma), \bm{\phi})$ where $\bm{\mu}$ is the mean of the Gaussian model, $\Sigma$ is the covariance, and $\bm{\phi}$ is the probability of this Gaussian model. So we calculate the positioning $D_{j,z}\in D$ of a business with pricing $\bm{v}_j$ to an expenditure grade  with the normalized probability distribution of the business $j$ on the corresponding Gaussian model $z$.
\begin{equation}\label{eq:djz}
D_{j,z}=\frac{N(\bm{v}_{j};\mu _{z},\Sigma _{z})\cdot \phi _{z}}{\sum^{T}_{t=1}N(\bm{v}_{j};\mu _{t},\Sigma _{t})\cdot \phi _{t})}.
\end{equation}
Furthermore, we can consider users also have different sentiment on these expenditure grades. Here we define the weight matrix $W\in \mathbb{R}{^{m\times T}}$ to represent the sentiment on each expenditure grade for each user, which reflects user spending behaviors. We also call $W$ sentiment matrix in the following section.

The construction of $D$ and $W$ have many advantages. (1) We define the probability distribution matrix $D$ to represent the positioning of business to expenditure grades, not specific expenditure values, so it can eliminates the effect of expenditure scale. Moreover, multiple expenditure grades are statistically deduced from the whole expenditure range with the EM algorithm, and the positioning are probability values, not specific expenditure values. They are helpful to alleviate the effect of sparse and noisy expenditures. (2) The design of $D$ and $W$ not only complies with the rules observed from the data, but also reflects the business pricing models and user spending behaviors, which are also validated in the experiments in Section 4.5.

We then leverage $W$ and $D$ to adjust user and business latent factors in the matrix factorization as follows,
\begin{equation}\label{eq:ob3}
\begin{aligned}
\underset{P,Q,W}{\min}&\ \frac{1}{2}\parallel I\odot (R-PQ^{\top}-WD^{\top}) \parallel_{F}^{2}\\&+\gamma (\parallel P \parallel_{F}^{2}+\parallel Q \parallel_{F}^{2})+\beta\parallel W \parallel_{F}^{2}.
\end{aligned}
\end{equation}
In this way, we can integrate the all three rules in the optimization objective. The optimization objective is similar to that in \cite{LIAN2014}. In fact, the augmented matrix $[P, W]$ and $[Q,D]$ can be considered as the extended latent factors of users and businesses, respectively.  The extended latent factors of users and businesses not only stem from the rating information (i.e., $P$ and $Q$) but also consider the expenditure information (i.e., $W$ and $D$). However, different from \cite{LIAN2014},  the $D$ is directly deduced from expenditure matrix, which implies less parameters to be learned in EARP. Moreover, they integrate totally different information with different frameworks.

In order to solve the above objective function, we can rewrite Eq. (\ref{eq:ob3}) as follows,
\begin{equation} \label{eq:ob3-1}
\begin{split}
& \mathcal{L}(P, Q, W)=\frac{1}{2}\sum_{i=1}^{m}\sum_{j=1}^{n}I_{i,j}(R_{i,j}-P_{i}Q_{j}^{\top}-W_{i}D_{j}^{\top})\\
& +\frac{1}{\gamma}\sum_{i=1}^{m}\parallel P_{i} \parallel_{F}^{2}+\frac{1}{\gamma}\sum_{j=1}^{n}\parallel Q_{j} \parallel_{F}^{2}+\frac{1}{\beta }\sum_{i=1}^{m}\parallel W_{i} \parallel_{F}^{2}.
\end{split}
\end{equation}
Gradient descent method can be used to solve this problem. There are three variables, including the user latent factor $P$, the business latent factor $Q$ and the weight matrix $W$. $P_{i}$ is user $i$'s latent factor, $Q_{j}$ is business $j$'s latent factor and $W_{i}$ is user $i$'s weight vector. A local minimum of the objective function given by Eq. (\ref{eq:ob3-1}) can be obtained by applying gradient descent to $P_{i}$, $Q_{j}$ and $W_{i}$ as
\begin{equation}\label{eq:123}
\begin{split}
& \frac{\partial \mathcal{L}}{\partial P_{i}}=\sum_{j=1}^{n}I_{i,j}(P_{i}Q_{j}^{\top}+W_{i}D_{j}^{\top}-R_{i,j})\cdot Q_{j}^{\top}+\gamma P_{i} \\
& \frac{\partial \mathcal{L}}{\partial Q_{j}}=\sum_{i=1}^{m}I_{i,j}(P_{i}Q_{j}^{\top}+W_{i}D_{j}^{\top}-R_{i,j})\cdot P_{i}+\gamma Q_{j} \\
& \frac{\partial \mathcal{L}}{\partial W_{i}}=\sum_{j=1}^{n}I_{i,j}(P_{i}Q_{j}^{\top}+W_{i}D_{j}^{\top}-R_{i,j})\cdot D_{j}^{\top}+\beta W_{i}.
\end{split}
\end{equation}
The detailed framework of EARP is shown in Algorithm 2.

After obtaining the final solution of $P$, $Q$, and $W$, we get the estimated preference matrix for our proposed EARP-M model as follows
\begin{equation}
\widetilde{R}=PQ^{T}+WD^{T}.
\end{equation}
\renewcommand{\algorithmicrequire}{\textbf{Input:}}
\renewcommand{\algorithmicensure}{\textbf{Output:}}
\begin{algorithm}[t]
    \caption{Framework of EM algorithm for expenditure clustering}\label{Alg:EM}
    \begin{algorithmic}[1]
    \REQUIRE ~~\\
        $C$: Expenditure matrix\\
        $T$: Number of expenditure grades
    \ENSURE ~~\\
        $\bm{\mu}$: Means of the Gaussian\\
        $\Sigma$: Covariance of the Gaussian\\
        $\bm{\phi}$:  Probability of of the multinomial distribution
    \REPEAT
        \STATE E-step: calculate the posterior probability of multinomial distribution
        \FOR {each expenditure $e_{i}\in C$ and grade $z$}
            \STATE set $\omega_{z}^{i}:=\frac{N(e_{i};\mu _{z},\Sigma _{z})\cdot \phi _{z}}{\sum^{T}_{t=1}N(e_{i};\mu _{t},\Sigma _{t})\cdot \phi _{t}}$
        \ENDFOR
        \STATE M-step: update the parameters:
        \STATE $\phi _{z} := \frac{1}{m\times n}\sum_{i=1}^{m\times n}\omega _{z}^{i}$
        \STATE $\mu _{z} := \frac{\sum_{i=1}^{m\times n}\omega _{z}^{i}e_{i}}{\sum_{i=1}^{m\times n}\omega _{z}^{i}}$
        \STATE $\Sigma _{z} := \frac{\sum_{i=1}^{m\times n}\omega _{z}^{i}(e_{i}-\mu _{z})(e_{i}-\mu _{z})^{\top}}{\sum_{i=1}^{m\times n}\omega _{z}^{i}}$
    \UNTIL{convergence}
    \end{algorithmic}
\end{algorithm}

\begin{algorithm}[ht]
    \caption{Framework of EARP-M}\label{Alg:EARP}
    \begin{algorithmic}[1]
    \REQUIRE ~~\\
        $R$: Rating matrix\\
        $C$: Expenditure matrix\\
        $T$: Number of expenditure grades\\
        $\alpha _{0}$: Step size for updating parameters for $P$ and $Q$\\
        $\alpha _{1}$: Step size for updating parameters for $W$\\
        $\epsilon $: Convergence tolerance\\
    \ENSURE ~~\\
        $P$: User latent factor  \\
        $Q$: Business latent factor \\
        $W$: Sentiment matrix of users to expenditure grades

      \STATE Calculate business pricing $\bm{v}$ with Eq. (\ref{eq:vj})
     \STATE Calculate mixture Gaussian distributions with Alg. 1
     \STATE Calculate positioning matrix $D$ with Eq. (\ref{eq:djz})

    \STATE Randomly initialize $P>0$, $Q>0$, $W>0$
    \REPEAT
        \STATE Set $P_{old}:=P$, $Q_{old}:=Q$, $W_{old}:=W$\\
        \STATE Calculate $\frac{\partial \mathcal{L}}{\partial P}$, $\frac{\partial \mathcal{L}}{\partial Q}$, $\frac{\partial \mathcal{L}}{\partial W}$ with Eq. (\ref{eq:123})\\
        \STATE $P :=P-\alpha _{0}\frac{\partial \mathcal{L}}{\partial P}$\\
        \STATE $Q :=Q-\alpha _{0}\frac{\partial \mathcal{L}}{\partial Q}$\\
        \STATE $W :=W-\alpha _{1}\frac{\partial \mathcal{L}}{\partial W}$\\
    \UNTIL {$\mid P-P _{old} \mid+\mid Q-Q_{old} \mid+\mid W-W_{old} \mid < \epsilon $}
    \end{algorithmic}
\end{algorithm}

%
\subsection {Discussion}
We propose three models to sequentially integrate three rules observed in Section~\ref{sec:ob}. The preceding model can be considered as the special case of the successive model. The EARP-M model (see Eq. (\ref{eq:ob3})) converts to the EARP-U model (see Eq. (\ref{eq:ob2})), if the number of expenditure grades is 1. And the EARP-U model converts to the EARP-E model (see Eq. (\ref{eq:ob1})), when all weights  in $\bm{w}$ are equal. So the successive model is able to more comprehensively integrate the correlation of expenditures and rating scores, and it hopefully achieves better performances.

In the proposed expenditure term, we put forward two novel matrices: $D$ meaning the positioning of business to expenditure grades and $W$ representing the sentiment of users to expenditure grades. The positioning matrix $D$ is directly deduced through the expenditure matrix $C$ with the EM algorithm (see Algorithm 1), and the weight matrix $W$ is iteratively learned together with $P$ and $Q$ (see Algorithm 2). Moreover, these two proposed matrices $D$ and $W$ have the obvious physical significance that insightfully reflect the characteristics of businesses and users. The positioning matrix $D$ reflects the characteristics of expenditure distributions and reveals the business pricing models. 

Let's analyze the time complexity of the algorithm. The main computation lies in the EM algorithm calculating mixture Gaussian distributions (see Algorithm~\ref{Alg:EM}) and the gradient descent method iteratively optimizing $P$, $Q$, and $W$ (see Lines 5-11 in Algorithm~\ref{Alg:EARP}). Let $N$ be the number of nonzero entries in $R$, $k$ be the number of latent factors, and $T$ be the number of expenditure grades. The computational complexity of the EM algorithm for estimating mixture Gaussian model in each iteration is $O(N \times T)$. The main computation of the gradient descent method is to evaluate the objective function and its gradients. The computational complexity of estimating Eq. (\ref{eq:ob3}) is $O(N\times (T+k))$. The computational complexity for gradients $\frac{\partial \mathcal{L}}{\partial P_{i}}$, $\frac{\partial \mathcal{L}}{\partial Q_{j}}$ is $O(N\times k)$  and $\frac{\partial \mathcal{L}}{\partial W_{i}}$ is $O(N\times T)$. Thus, the computational complexity for each iteration in Algorithm~\ref{Alg:EARP} is $O(N\times (T+k))$.  In summary, the computational complexity of the algorithm is $O(N\times (T+k))$, linear with respect to the number of rating scores in matrix $R$.

\begin{table*}[!th]
  \centering \scriptsize
  \caption{Experimental Results of Effectiveness on RMSE Measure}
  {
    \begin{tabular}{@{}c|c|c|c|c|c|c|c|c||c|c|c@{}}
    \hline
    \textbf{Data set} & \textbf{Training Set} & \textbf{UserMean} & \textbf{ItemMean} &\textbf{PMF} & \textbf{LLORMA} & \textbf{cPMF} & \textbf{FM} & \textbf{CoMF}& \textbf{EARP-E} & \textbf{EARP-U} & \textbf{EARP-M} \\
    \hline
    \hline
    \multirow{4}[0]{*}{\textbf{BJ}}
    & {\textbf{60\%}}    & 0.9005& 0.9224& 0.8853& 0.8696& 0.8920& 0.8638& 0.8835& 0.8657& 0.8589&\textbf{0.8449}\\
          \cline{2-12}
    & {\textbf{70\%}}    & 0.8970& 0.9206& 0.8829& 0.8651& 0.8907& 0.8597& 0.8807& 0.8661& 0.8565&\textbf{0.8430}\\
          \cline{2-12}
    & {\textbf{80\%}}    & 0.8921& 0.9173& 0.8767& 0.8602& 0.8818& 0.8563& 0.8751& 0.8628& 0.8531&\textbf{0.8384}\\
          \cline{2-12}
    & {\textbf{90\%}}    & 0.8878& 0.9127& 0.8698& 0.8534& 0.8711& 0.8528& 0.8703& 0.8590& 0.8486&\textbf{0.8359}\\
          \cline{2-12}
    \hline
    \hline
    \multirow{4}[0]{*}{\textbf{SH}}
    & {\textbf{60\%}}   & 0.8963& 0.8965& 0.8710& 0.8551& 0.8695& 0.8524& 0.8680& 0.8549& 0.8475&\textbf{0.8342} \\
          \cline{2-12}
    & {\textbf{70\%}}   & 0.8918& 0.8934& 0.8632& 0.8499& 0.8644& 0.8474& 0.8657& 0.8524& 0.8441&\textbf{0.8313} \\
          \cline{2-12}
    & {\textbf{80\%}}   & 0.8860& 0.8907& 0.8557& 0.8440& 0.8533& 0.8423& 0.8525& 0.8484& 0.8391&\textbf{0.8263} \\
          \cline{2-12}
    & {\textbf{90\%}}   & 0.8835& 0.8909& 0.8510& 0.8424& 0.8489& 0.8416& 0.8506& 0.8483& 0.8377&\textbf{0.8258} \\
         \cline{2-12}
    \hline
    \hline
    \multirow{4}[0]{*}{\textbf{GZ}}
    & {\textbf{60\%}}   & 0.8644& 0.9149& 0.8527& 0.8494& 0.8542& 0.8470& 0.8476& 0.8428& 0.8404&\textbf{0.8237} \\
           \cline{2-12}
    & {\textbf{70\%}}   & 0.8616& 0.9089& 0.8504& 0.8429& 0.8518& 0.8431& 0.8455& 0.8399& 0.8370&\textbf{0.8203} \\
           \cline{2-12}
    & {\textbf{80\%}}   & 0.8574& 0.9076& 0.8489& 0.8384& 0.8504& 0.8409& 0.8436& 0.8386& 0.8345&\textbf{0.8180} \\
           \cline{2-12}
    & {\textbf{90\%}}    & 0.8583& 0.9070& 0.8514& 0.8395& 0.8526& 0.8380& 0.8465& 0.8418& 0.8372&\textbf{0.8196}\\
           \cline{2-12}
    \hline
    \hline
    \multirow{4}[0]{*}{\textbf{HZ}}
    & {\textbf{60\%}}   & 0.9460& 0.9767& 0.9279& 0.9288& 0.9338& 0.9168& 0.9230& 0.9093& 0.9024&\textbf{0.8890} \\
           \cline{2-12}
    & {\textbf{70\%}}   & 0.9442& 0.9741& 0.9262& 0.9229& 0.9328& 0.9094& 0.9221& 0.9086& 0.9003&\textbf{0.8867} \\
           \cline{2-12}
    & {\textbf{80\%}}   & 0.9418& 0.9700& 0.9242& 0.9178& 0.9285& 0.9051& 0.9212& 0.9081& 0.8979&\textbf{0.8847} \\
           \cline{2-12}
    & {\textbf{90\%}}    & 0.9372& 0.9654& 0.9174& 0.9129& 0.9221& 0.9000& 0.9172& 0.9040& 0.8935&\textbf{0.8796}\\
           \cline{2-12}
    \hline
    \hline
    \multirow{4}[0]{*}{\textbf{CD}}
    &{\textbf{60\%}}   & 0.8963& 0.9561& 0.8829& 0.8858& 0.8848& 0.8815& 0.8766& 0.8724& 0.8682&\textbf{0.8529} \\
           \cline{2-12}
    & {\textbf{70\%}}    & 0.8904& 0.9511& 0.8816& 0.8812& 0.8838& 0.8816& 0.8753& 0.8712& 0.8652&\textbf{0.8493}\\
           \cline{2-12}
    & {\textbf{80\%}}   & 0.8877& 0.9439& 0.8761& 0.8743& 0.8896& 0.8770& 0.8702& 0.8656& 0.8608&\textbf{0.8454} \\
           \cline{2-12}
    & {\textbf{90\%}}    & 0.8857& 0.9436& 0.8781& 0.8701& 0.8924& 0.8805& 0.8727& 0.8676& 0.8611&\textbf{0.8449}\\
           \cline{2-12}
    \hline

    \end{tabular}}%
  \label{tab:addlabel-3}%
\end{table*}%

\begin{table*}[!th]
  \centering  \scriptsize
  \caption{Experimental Results of Effectiveness on MAE Measure}
  {
    \begin{tabular}{@{}c|c|c|c|c|c|c|c|c||c|c|c@{}}
    \hline
    \textbf{Dataset} & \textbf{Training Set} & \textbf{UserMean} & \textbf{ItemMean} &\textbf{PMF} & \textbf{LLORMA} & \textbf{cPMF} & \textbf{FM} & \textbf{CoMF} & \textbf{EARP-E} & \textbf{EARP-U} & \textbf{EARP-M} \\
    \hline
    \hline
    \multirow{4}[0]{*}{\textbf{BJ}}
    & {\textbf{60\%}}  & 0.6788& 0.7099& 0.6734& 0.6547& 0.6740& 0.6744& 0.6697& 0.6747& 0.6574&\textbf{0.6452}\\
           \cline{2-12}
    & {\textbf{70\%}}  & 0.6764& 0.7079& 0.6720& 0.6512& 0.6729& 0.6695& 0.6682& 0.6760& 0.6555&\textbf{0.6438} \\
           \cline{2-12}
    & {\textbf{80\%}}   & 0.6732& 0.7058& 0.6680& 0.6474& 0.6670& 0.6667& 0.6645& 0.6644& 0.6530&\textbf{0.6402}\\
           \cline{2-12}
    & {\textbf{90\%}}   & 0.6711& 0.7027& 0.6643& 0.6429& 0.6603& 0.6643& 0.6621& 0.6681& 0.6503&\textbf{0.6384}\\
           \cline{2-12}
    \hline
    \hline
    \multirow{4}[0]{*}{\textbf{SH}}
    & {\textbf{60\%}}   & 0.6717& 0.6864& 0.6602& 0.6431& 0.6556& 0.6607& 0.6540& 0.6616& 0.6452&\textbf{0.6348}\\
           \cline{2-12}
    & {\textbf{70\%}}   & 0.6688& 0.6843& 0.6552& 0.6389& 0.6516& 0.6560& 0.6523& 0.6573& 0.6427&\textbf{0.6325}\\
           \cline{2-12}
    & {\textbf{80\%}}   & 0.6652& 0.6824& 0.6509& 0.6349& 0.6444& 0.6515& 0.6445& 0.6534& 0.6393&\textbf{0.6292}\\
           \cline{2-12}
    & {\textbf{90\%}}   & 0.6633& 0.6822& 0.6474& 0.6332& 0.6413& 0.6504& 0.6429& 0.6552& 0.6378&\textbf{0.6279}\\
           \cline{2-12}
    \hline
    \hline
    \multirow{4}[0]{*}{\textbf{GZ}}
    & {\textbf{60\%}}   & 0.6583& 0.7063& 0.6629& 0.6406& 0.6560& 0.6717& 0.6561& 0.6625& 0.6501&\textbf{0.6344}\\
           \cline{2-12}
    & {\textbf{70\%}}   & 0.6559& 0.7013& 0.6614& 0.6352& 0.6542& 0.6679& 0.6546& 0.6536& 0.6476&\textbf{0.6314}\\
           \cline{2-12}
    & {\textbf{80\%}}   & 0.6536& 0.7019& 0.6618& 0.6333& 0.6543& 0.6665& 0.6547& 0.6563& 0.6467&\textbf{0.6303}\\
           \cline{2-12}
    & {\textbf{90\%}}   & 0.6541& 0.7014& 0.6630& 0.6336& 0.6548& 0.6611& 0.6560& 0.6599& 0.6483&\textbf{0.6305}\\
           \cline{2-12}
    \hline
    \hline
    \multirow{4}[0]{*}{\textbf{HZ}}
    & {\textbf{60\%}}   & 0.7145& 0.7597& 0.7204& 0.6996& 0.7170& 0.7325& 0.7116& 0.7127& 0.6974&\textbf{0.6832}\\
           \cline{2-12}
    & {\textbf{70\%}}   & 0.7128& 0.7586& 0.7192& 0.6955& 0.7158& 0.7232& 0.7110& 0.7118& 0.6959&\textbf{0.6816}\\
           \cline{2-12}
    & {\textbf{80\%}}   & 0.7102& 0.7549& 0.7172& 0.6914& 0.7121& 0.7175& 0.7095& 0.7058& 0.6941&\textbf{0.6790}\\
           \cline{2-12}
    & {\textbf{90\%}}   & 0.7065& 0.7531& 0.7128& 0.6884& 0.7078& 0.7136& 0.7071& 0.7032& 0.6913&\textbf{0.6761}\\
           \cline{2-12}
    \hline
    \hline
    \multirow{4}[0]{*}{\textbf{CD}}
    & {\textbf{60\%}}   & 0.6834& 0.7407& 0.6875& 0.6684& 0.6811& 0.7064& 0.6796& 0.6811& 0.6730&\textbf{0.6585}\\
           \cline{2-12}
    & {\textbf{70\%}}   & 0.6790& 0.7372& 0.6871& 0.6637& 0.6808& 0.7066& 0.6792& 0.6784& 0.6708&\textbf{0.6555}\\
           \cline{2-12}
    & {\textbf{80\%}}   & 0.6776& 0.7322& 0.6834& 0.6596& 0.6761& 0.6919& 0.6757& 0.6774& 0.6682&\textbf{0.6529}\\
           \cline{2-12}
    & {\textbf{90\%}}   & 0.6763& 0.7320& 0.6860& 0.6580& 0.6765& 0.6915& 0.6785& 0.6789& 0.6696&\textbf{0.6532}\\
           \cline{2-12}
    \hline

    \end{tabular}}%
  \label{tab:addlabel-4}%
\end{table*}%

\section{Experiments}
In this section, extensive experiments on the above five real data sets illustrate the traits of EARP from four aspects. We first validate the effectiveness of EARP through comparing it with other methods.  Then we show the effect of the number of expenditure grades on performances. Furthermore, we thoroughly explore the characteristics of EARP on the meanings of the $D$ and $W$ matrix. Finally, we validate the sensitivity of EARP to the missing expenditure information.

\subsection{Metrics}
We use two widely used metrics, Root Mean Square Error (RMSE) and Mean Absolute Error (MAE), to measure the rating prediction quantity. RMSE is defined as:
\begin{equation}\label{eq:rmse}
\text{RMSE}=\sqrt{\frac{\sum_{(p,q)\in R}{(r_{p,q}-\hat{r}_{p,q})^2}}{|R|}},
\end{equation}
where $r_{p,q}$ denotes the real rating user $p$ gave to item $q$ and $\hat{r}_{p,g}$ denotes the predicted rating. $R$ denotes the whole rating set. MAE is defined as:
\begin{equation}\label{eq:mae}
\text{MAE}=\frac{\sum_{(p,q)\in R}{|r_{p,q}-\hat{r}_{p,q}|}}{|R|}.
\end{equation}

Smaller values of MAE or RMSE mean better performances. RMSE is more sensitive and stricter to the large error in comparison with MAE because of the square penalty.

\subsection{Comparison Methods}
In order to show the effectiveness of the proposed EARP, we compare three variations of EARP with representative rating predication methods. The first four methods only utilize the rating information. The middle three methods (i.e., cPMF, FM, and CoMF) utilize the rating and expenditure information.
\begin{itemize}
\item UserMean. This method uses the mean value of every user to predict the missing values.
\item ItemMean. This method utilizes the mean value of every item to predict the missing values.
\item PMF~\cite{MNIH2007}: It is the basic matrix factorization method only using user-item rating matrix for recommendations.
\item LLORMA~\cite{LEE2013}: It is the newest matrix factorization method based on local low-rank assumption. We use it on rating matrix for recommendations.
\item cPMF~\cite{GE2011}: It is a cost-aware travel tour recommendation method, which utilizes the business pricings as cost.
\item FM~\cite{rendle2012factorization}: It is a generic approach that allows to mimic most factorization model by feature engineering. We use it on rating matrix and treat the business pricings as features.
\item CoMF: It is a collective matrix factorization method that factorizes rating score and expenditure matrix simultaneously and shares a common user latent factor.
\item EARP-E: It just integrates the expenditure correlation rule.
\item EARP-U: It integrates the expenditure correlation and user spending correlation rules.
\item EARP-M: It is the standard version of EARP, which integrates all three observed rules.
\end{itemize}

For EARP, we set the number of expenditure grades to be 5, as suggested in the following experiments. The regularization coefficient $\gamma$  is set as 0.08, and the regularization coefficient $\beta$ is set as 0.1. The parameters in other methods are set with the best performances on these datasets by cross validation. Note that, for those methods (i.e., cPMF, FM, and CoMF) utilizing expenditure information, we employ the same feature processing in EARP to avoid scale mismatch of rating and expenditure.

\subsection{Effectiveness Study}

For each data set, we randomly select training data with different ratios, varying from 60\% to 90\% with the interval of 10\%, from user-item rating matrix, and the rest as test data. The random selection is repeated 10 times independently and the average results are reported in Tables~\ref{tab:addlabel-3} and~\ref{tab:addlabel-4}.


From the results, we have the following observations:
 \begin{itemize}
   \item The three versions of EARP outperform other approaches in most conditions. Particularly, EARP-M always achieves the best performances in all conditions. It confirms that the expenditure factor is able to improve the accuracy of rating scores. In addition, due to the lowest sparsity of HZ data set, all methods have poor performances on this data set.
   \item Due to ignoring the expenditure information, the methods only utilizing rating score information (e.g., LLORMA and PMF) have poor performances. It validates that the expenditure information is really important for recommendation.
   \item The three methods utilizing both rating score and expenditure information (i.e., cPMF, FM, and CoMF) also perform worse than EARP in most conditions. Although these methods can integrate expenditure information, they simply utilize expenditure information as features, without considering the unique characteristics of expenditure information, so they achieve worse performances than EARP.  Because of directly factorizing expenditure matrix, CoMF introduces much noise, which leads to its poor performances. The proposed EARP model not only makes full use of the correlations of expenditures and rating scores, but also  has a good mechanism to integrate these correlations. So EARP has the best performances on most conditions.
   \item We can also find that three versions of EARP have different performances. EARP-M always provides the best performances, while EARP-U outperforms EARP-E. The performance behaviors of these three versions of EARP confirm our observations and assumptions. EARP-E confirms that the expenditure term can improve the rating prediction performance, and EARP-U shows the effectiveness of personalized weights of users. The best performances of EARP-M show the benefits from considering multiple expenditure grades.
 \end{itemize}

\begin{figure*}
\centering \label{fig:weight}
\subfigure[Ave. weights of all users on expenditure grades]{
  \includegraphics[width=4cm]{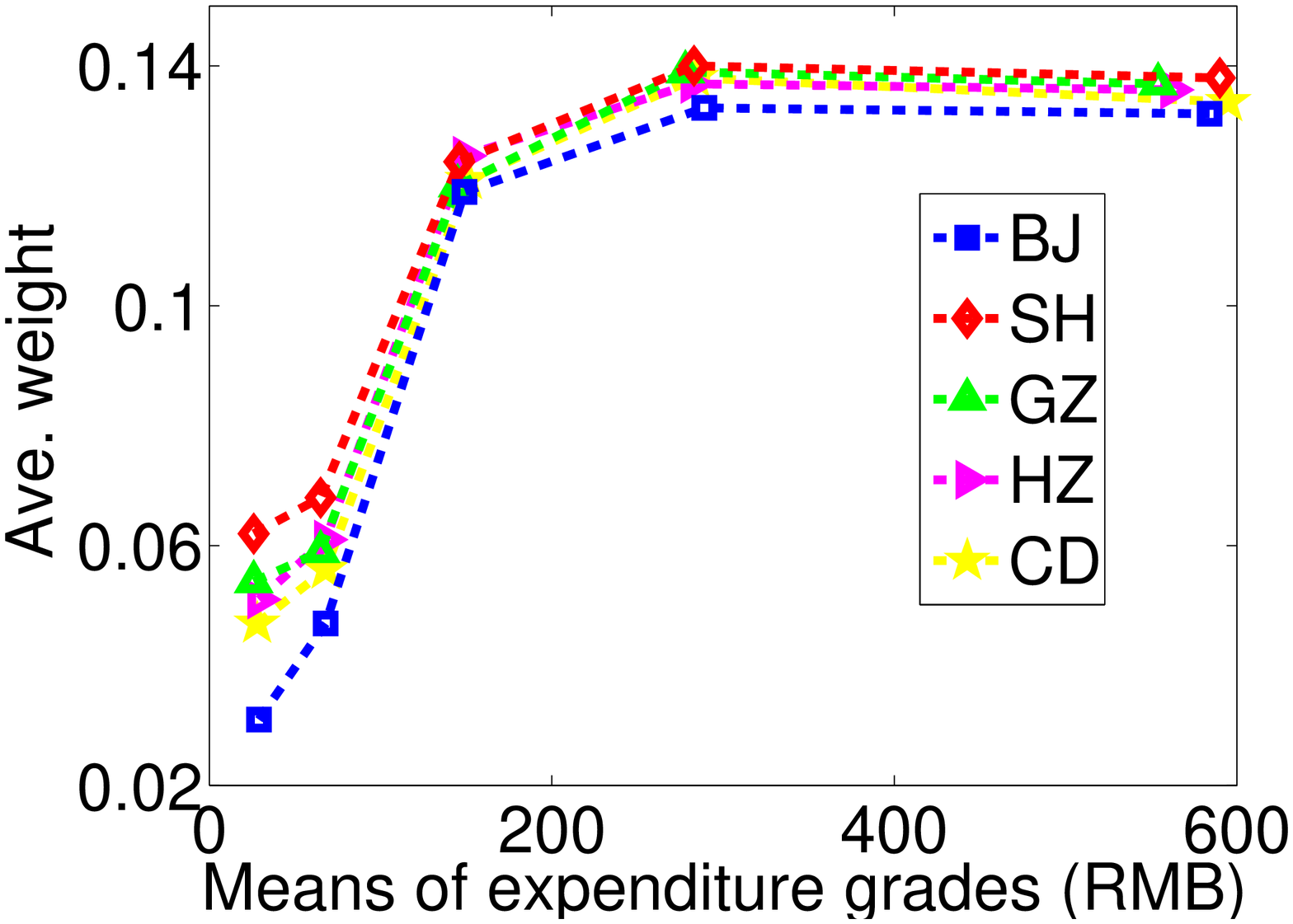} \label{fig:weight-a}
}
\subfigure[Ave. weights of specific users on expenditure grades]{
  \includegraphics[width=4cm]{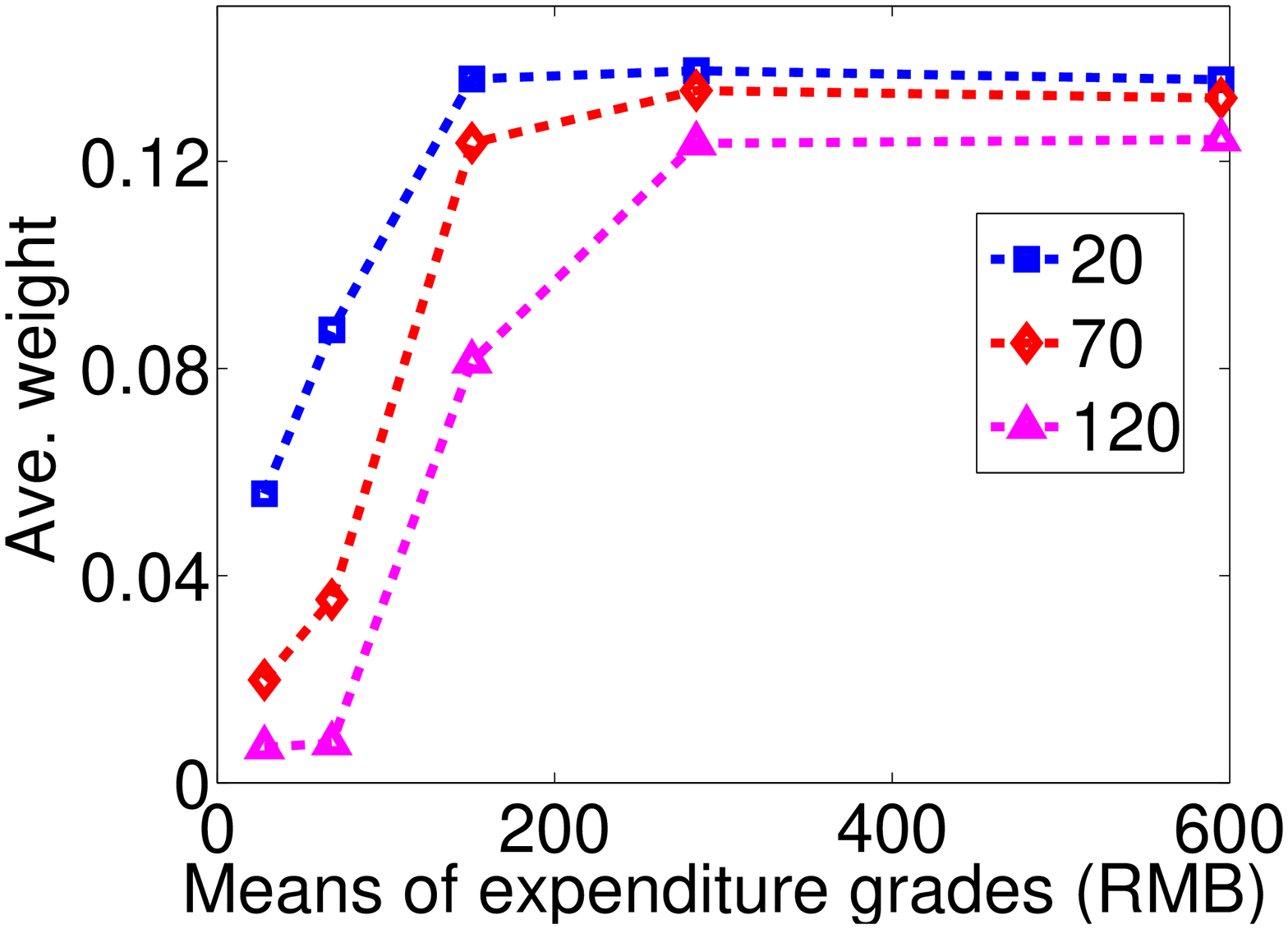} \label{fig:weight-b}
}
\subfigure[Ave. weights of users with different spendings ]{
  \includegraphics[width=4cm]{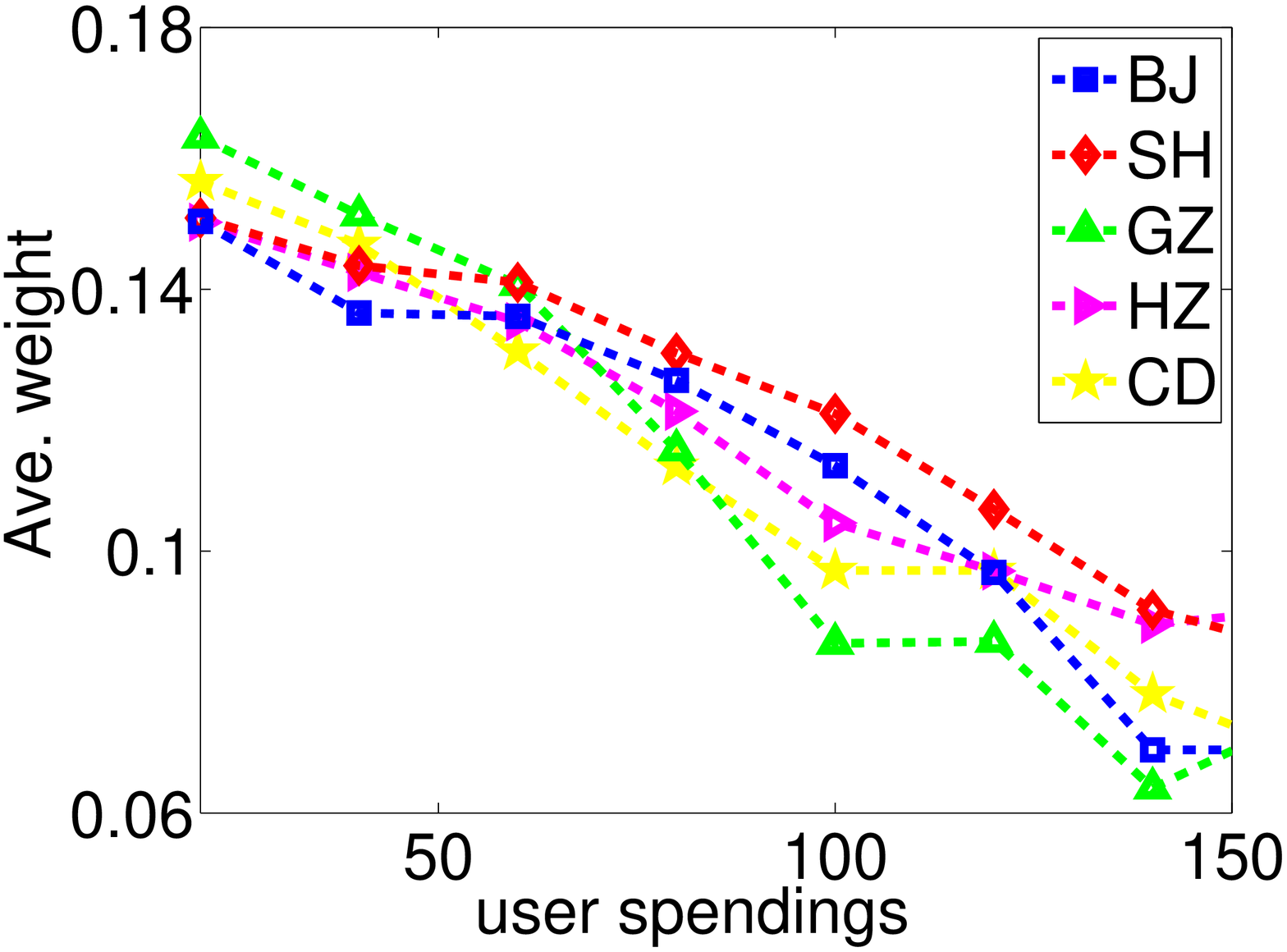} \label{fig:weight-c}
}
\subfigure[Positioning of businesses on expenditure grades]{
  \includegraphics[width=4cm]{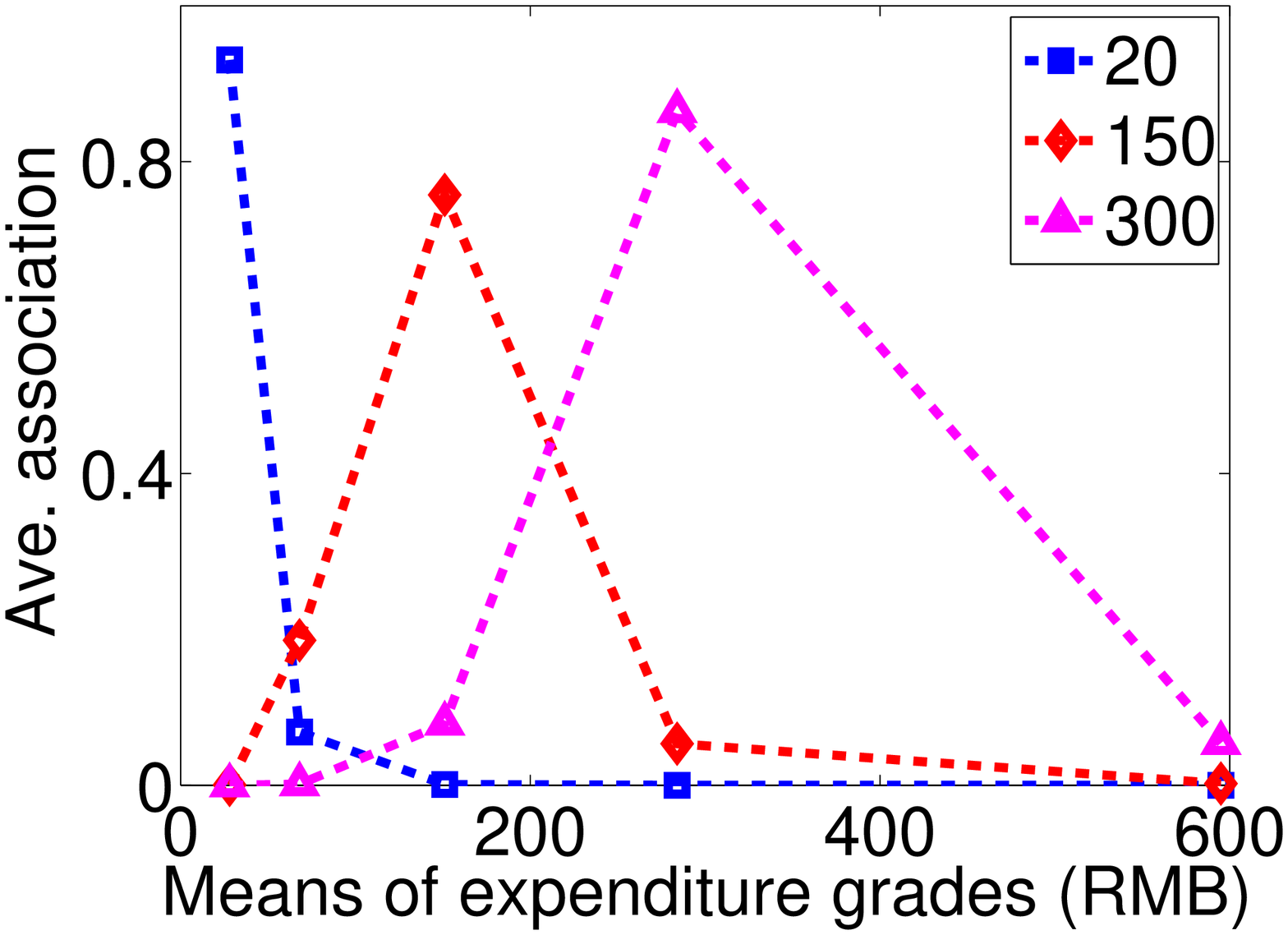} \label{fig:weight-d}
}
\caption{Statistics of sentiment and positioning matrices.  }
\end{figure*}

\subsection{Meanings Study of Sentiment and Positioning Matrices}
In this section, we investigate the meanings of the sentiment and positioning matrices learned by EARP-M through a case study. Based on the matrices learned by  EARP-M with $80\%$ training data in the above experiments, we do the following four experiments.

In the first experiment, we observe the average weights of users (i.e., $\frac{1}{m}\sum_{i}W_{i,t}$, $t\in [1, T]$, and $T = 5$) on five expenditure grades. We first show the average weights of users on five data sets in Fig.~\ref{fig:weight-a}. Generally, it is obvious that users' weights are higher on higher expenditure grades. According to  the Observation 1 that the users with higher expenditures tend to give higher rating scores, we can find that the weight reflects the sentiment of users to expenditures. That is, the higher weight on higher expenditure grades means that users are more sensitive to the higher expenditures.

Furthermore, in the second experiment, we intensively observe the average weights of users with a specific spending on expenditure grades. Concretely, we select three types of users: users' average spendings are around 20, 70, and 120, which represents low-consumption, middle-consumption, and high-consumption users, respectively. And then we calculate the average weights of each type of users on five expenditure grades. The results are shown in Fig.~\ref{fig:weight-b}. Besides the similar observation as in Fig.~\ref{fig:weight-a}, we also find that the higher spending users have lower weights. That is to say, a high spending user is not very sensitive to his expenditure. This is reasonable. When a high spending user receives service in a business, he usually cares more about the service, and notice less about the expenditure.

In the third experiment, we observe the average weights of users with different spendings on a specific expenditure grade. We divide user spendings into 7 regions with the interval of 20, and calculate the average weights of users in each region. Fig.~\ref{fig:weight-c} shows the results on a specific expenditure grade, i.e.,  RMB 149. It is clear that higher spending users has lower weights. It confirms the above experimental conclusion again. That is,  a high spending user is less sensitive to the same expenditure than a low spending user. It also complies with the Observation 2 that the rating scores are positively correlated to the differences between the actual expenditure and user spending.

Finally, we explore the positioning of businesses on expenditure grades (i.e., $D$ matrix). We show the possibility values of three types of businesses: business pricing around 20 (i.e., low-grade business), 150 (i.e., middle-grade business), and 300 (i.e., high-grade business). The results are shown in Fig.~\ref{fig:weight-d}. It is very clear that the businesses have obvious high possibility on the closest expenditure grade to their business pricings, which correctly reflects business pricing models. Taking a middle-grade business as an example (i.e., 150), its main expenditures are around 150, along with a small amount of expenditures on low-grade (e.g., 20) or high-grade (e.g., 300) consumptions.

In all, the positioning matrix $D$ discovers pricing model of a business. Moreover, the sentiment matrix $W$ reveals user spending behaviors. That is, users are more sensitive to a high expenditure, while a high spending user has less sensitivity to an expenditure than a low spending user. Through the expenditure term, the EARP method provides insight to user spending behaviors and business pricing models, which is not addressed by those baselines.

\begin{figure}
\centering
\subfigure[RMSE]{
  \includegraphics[width=3.9cm]{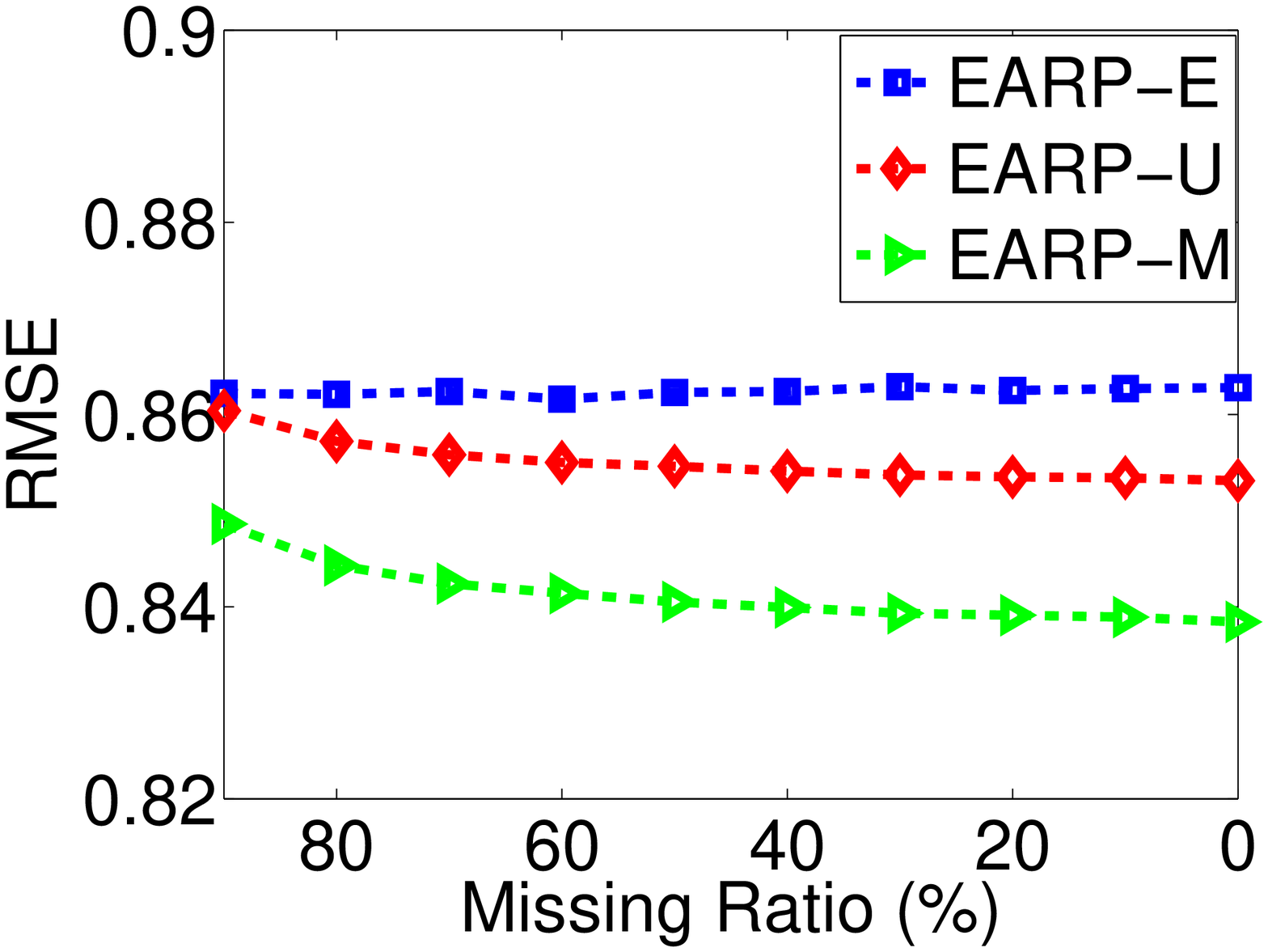}
}
\subfigure[MAE]{\label{fig:fft:a}
  \includegraphics[width=3.9cm]{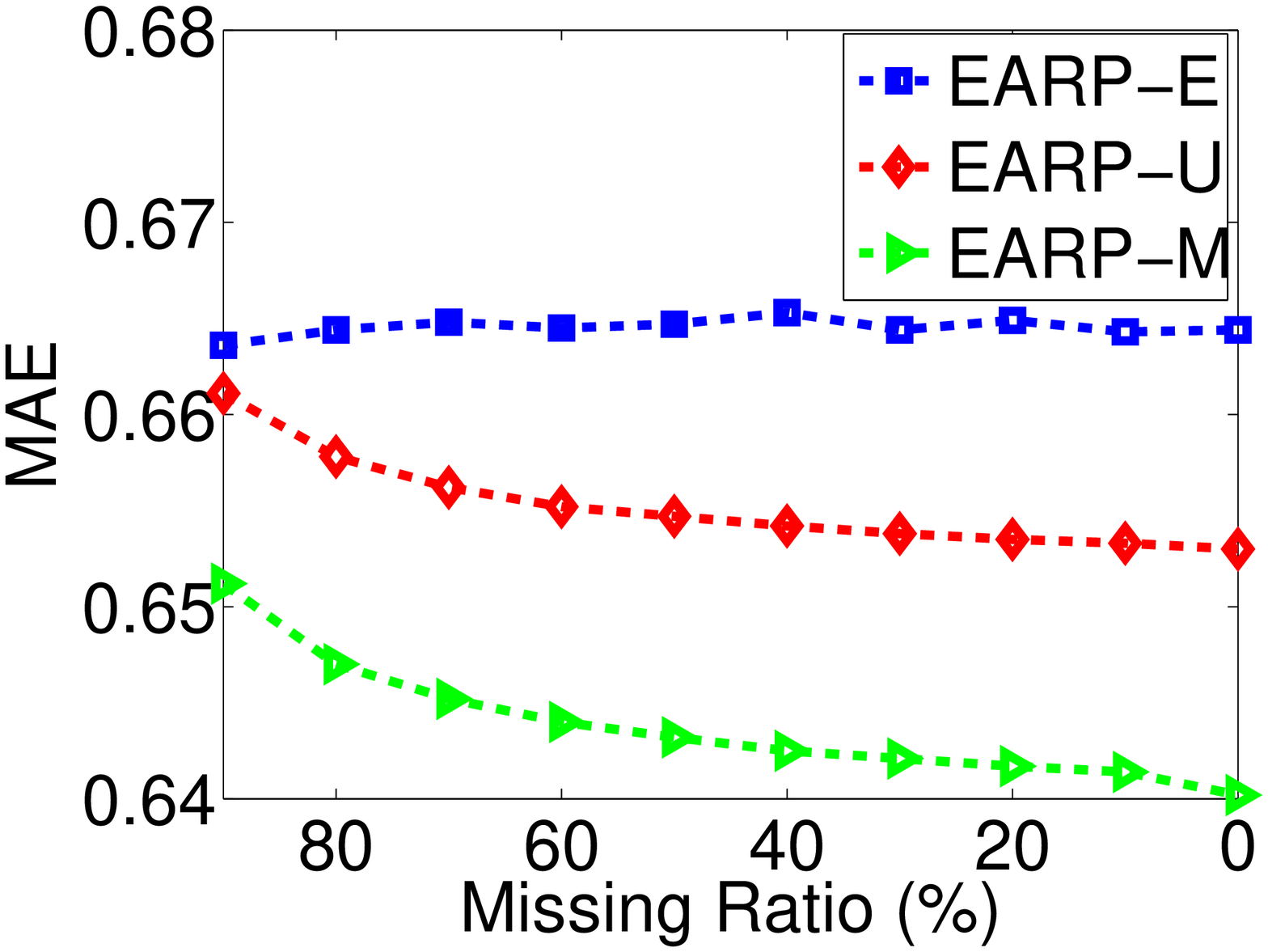}
}
\caption{Performances of EARP on different ratio of missing expenditure.  }\label{fig:missingdata}
\end{figure}
\subsection{Sensitivity Study of Missing Expenditure Information}

In our model, we assume that each rating score of users on items has a corresponding expenditure. However, the expenditure information may be missing in real applications. In this section, we validate the effectiveness of EARP under different conditions of missing expenditure information. In BJ, we randomly drop the expenditure data with ratios from 10\% to 90\% with the  interval of 10\%. Then we run three versions of EARP on these data sets with 80\% training data. The results are shown in Fig.~\ref{fig:missingdata}. Generally, the performances of EARP degrade when more expenditure information is missing. However, three versions of EARP all are tolerant to missing expenditure data, since their performances do not degrade a lot for more missing expenditure data. Taking the EARP-M as an example, its performance degradation is no more than 1\% when the missing expenditure data is more than 80\%. In addition, we can see that EARP-E is not sensitive to the missing expenditure data. The reason is that all users share the same weight $w$ in this model, and EARP-E can learn weights from existing expenditure data.

\section{Conclusion}
In this paper, we first study the expenditure aware rating prediction problem and propose a novel solution EARP. Through analyzing real data from a well-known social media platform, we find some interesting rules between rating scores and expenditures. Utilizing these rules, we design EARP models based on low-rank matrix factorization. Extensive experiments on five real data sets validate the effectiveness of EARP.



%
\bibliographystyle{abbrv}
\bibliography{sigproc}

\end{document}